\documentclass[11pt]{article}

\usepackage[margin=1in]{geometry}
\usepackage[T1]{fontenc}
\usepackage{graphicx}
\usepackage{amsmath}
\usepackage{amssymb}
\usepackage{amsfonts}
\usepackage{bm}
\usepackage{booktabs}
\usepackage{multirow}
\usepackage{array}
\usepackage{caption}
\usepackage{subcaption}
\usepackage{enumitem}
\usepackage{gensymb}
\usepackage{cite}
\usepackage{hyperref}
\usepackage[nameinlink,noabbrev]{cleveref}

\hypersetup{
  colorlinks=true,
  linkcolor=blue,
  citecolor=blue,
  urlcolor=blue
}

\title{Experimental Validation of Combined Imaging and Vibration Mitigation for High-Altitude Platforms}

\author{%
\parbox{0.94\textwidth}{\centering
Ákos Radványi, Anita Keszler, Dániel Balogh, Béla Takarics, \textit{Member, IEEE}, \\
András Majdik, \textit{Member, IEEE}, Andrej Tokarjev, Gábor Kovács, \\
Tamás Szirányi, \textit{Senior Member, IEEE}, Bálint Vanek, \textit{Member, IEEE}\\[0.75em]
\small Á. Radványi, D. Balogh, B. Takarics, B. Vanek are with the Systems and Control Laboratory of HUN-REN SZTAKI, Budapest, Hungary.\\
\small A. Keszler, A. Majdik, A. Tokarjev, G. Kovács, T. Szirányi are with the Machine Perception Research Laboratory of HUN-REN SZTAKI, Budapest, Hungary.\\
\small Corresponding author: B. Takarics, e-mail: \texttt{takarics@sztaki.hu}}}

\date{}

\newcommand{\reportstatus}{%
  \begin{center}
    \begin{minipage}{0.94\textwidth}
      \small
      \textbf{Extended Technical Report.}
      This author-prepared report is based on a manuscript accepted for publication
      in \emph{IEEE/ASME Transactions on Mechatronics} and integrates additional
      supporting material supplied during peer review. The extended arrangement has
      not been separately peer reviewed or accepted by the journal.

      \medskip
      \textbf{Accepted for publication in IEEE/ASME Transactions on Mechatronics.}

      \medskip
      \textcopyright\ 2026 IEEE. Personal use of this material is permitted.
      Permission from IEEE must be obtained for all other uses, in any current or
      future media, including reprinting/republishing this material for advertising
      or promotional purposes, creating new collective works, for resale or
      redistribution to servers or lists, or reuse of any copyrighted component of
      this work in other works.
    \end{minipage}
  \end{center}
}

\begin{document}
\maketitle
\reportstatus

\begin{abstract}
The demand for continuous, high-resolution aerial monitoring is driving interest in High-Altitude Pseudo-Satellites (HAPS), which offer a cost-effective alternative to satellites with greater flexibility. While HAPS platforms provide valuable Earth observation potential, selecting an optical sensor requires balancing resolution and payload constraints, given their lightweight structures.
We propose a multidisciplinary concept to enhance the Earth observation capabilities of HAPS using a commercially available, weight-efficient distributed camera network, combined with advanced image processing and active vibration control. A super-resolution pipeline is introduced, where images from individual cameras are preprocessed and fed into a super-resolution algorithm. The low structural rigidity of HAPS platforms increases their sensitivity to low-frequency vibrations, which must be mitigated through active control to preserve imaging performance.
A nominal $\mathcal{H}_\infty$ mixed-sensitivity controller is designed to suppress resonance peaks at camera locations.
The methodology is experimentally validated on a simplified wing platform designed for real-time testing. Quantitative evaluation based on nine metrics demonstrates significant image quality improvements, highlighting the effectiveness of $\mathcal{H}_\infty$ control in mitigating low-frequency elastic modes and stabilizing the platform for super-resolution. These results show that lightweight commercial sensors, combined with advanced image processing and control, can deliver high-quality imaging for future HAPS remote sensing missions.
\end{abstract}


\section{Introduction} \label{sec:intro}


With recent advances in image processing and aircraft technology, there has been a growing need for autonomous and controllable aerial platforms that can bridge the gap between conventional aircraft and satellites. High Altitude Pseudo-Satellites (HAPS) are unmanned aerial vehicles operating at a fixed position relative to the Earth at altitudes of approximately 20 km or above, which is motivated by the fact that wind speed is less intense in this region, thus, less power is required to maintain their specified position. While traditional satellites are expensive to build and launch, HAPS technology offers the prospect of a more affordable and easily deployable alternative. In the present study, the term HAPS focuses on airplane-like platforms where the main challenges are the destructive vibrations coming from atmospheric winds and the limited payload capacity due to the lightweight, flexible structure. To enable long endurance, HAPS platforms use ultralight structures and high-aspect-ratio wings that maximize solar surface. For example, the AALTO Zephyr S has a 25 m wingspan, with only 75 kg structural mass, and it has demonstrated a record endurance of 67 days in 2025 \cite{zephyr}.


Traditionally, due to the flexibility of the wings, the HAPS imaging payload is a single, centrally mounted optical unit.
Since the slender, flexible structure of these platforms can accommodate only a few kilograms of onboard equipment, increasing the optical resolution by scaling conventional high-performance cameras is impractical from both weight and cost perspectives. Recent research in image processing suggests an alternative approach: rather than relying on a single high-performance sensor, multiple lightweight, commercially available cameras can be distributed along the wingspan, and their individual viewpoints can be fused through advanced image-processing techniques. Synchronized multi-camera systems have been successfully applied, for example, for capturing structural dynamics in civil engineering applications \cite{wang2022novel, wang2022completely}. In the context of aerial platforms, the optimal (rigid) arrangement of such camera systems has been analyzed in \cite{li2025pattern} using super-resolution methods.

Super-resolution (SR) is a well-studied research area with a broad range of applications, including remote sensing, surveillance, and medical imaging. SISR (Single-Image SR) techniques use a single low-resolution input, and exploit learned image priors or structural assumptions \cite{wang2020deep, dong2014learning, yu2016ultra, zhang2018residual, yang2019deep}. However, their performance is fundamentally limited by the lack of high-frequency information in the input image.
MISR (Multi-Image SR) approaches, on the other hand, combine multiple low-resolution inputs to reconstruct high-resolution images \cite{bhat2021deep, liu2022video, DeepLearning_SR_2024}.
 In terms of HAPS technology, the distributed camera configuration mounted along a flexible wing introduces characteristics of both multi-view and multi-frame MISR, motivating a hybrid pipeline, where multiple frames from each camera are fused to form the SR input. A central challenge in this setting lies in disturbance mitigation and geometric consistency among spatially separated sensors: structural vibrations and aeroelastic deformations continuously alter the relative camera poses, degrading registration accuracy and reconstruction quality.

Significant research has been conducted on the modeling and control of flexible unmanned aerial aircraft, where low-frequency elastic modes can strongly influence flight stability and structural integrity. Various control techniques — such as Linear Quadratic Regulator (LQR) \cite{Dillsaver2011Gust, Pereira2022Design}, $\mathcal{H}_\infty$ control \cite{Pusch2019Structured}, and boundary vibration control \cite{he2020dynamical} — have been proposed for gust load alleviation and flutter suppression, as demonstrated in the FliPASED H2020 project \cite{Patartics2021Structured}. Experimental research frameworks such as the UM/NAST system of the University of Michigan \cite{su2010nonlinear} and the flexible-wing test platform of the University of Bristol \cite{jayatilake2025nonlinear} have further advanced aeroelastic modeling and validation under laboratory conditions.
However, these studies primarily address flight dynamics and structural control, and no publicly documented HAPS initiative integrates a distributed imaging architecture with software-level SR reconstruction. Similarly, existing SR research with the sensor mounted on an aerial platform generally assumes a single-camera setup or rigid (or semi-rigid) camera configurations, and does not account for structural motion or time-varying inter-camera geometry \cite{suluhan2025hstr} \cite{li2023lswins_SR}. To conclude, there is currently no integrated framework that simultaneously considers active vibration suppression and multi-camera super-resolution on HAPS platforms under realistic structural deformation conditions.

To address these limitations, this paper introduces a system-level framework that combines lightweight distributed cameras, a super-resolution pipeline, and active vibration control (AVC) within a unique experimental setup. A scaled-down flexible-wing demonstrator was developed at HUN-REN SZTAKI, with two synchronized cameras and embedded actuation for AVC. This platform provides a safe and cost-effective environment for hardware-in-the-loop testing in the presence of expected structural deformation effects. The proposed framework focuses on the interaction of image processing and vibration mitigation: SR performance is evaluated in the presence of low-frequency oscillations, and the effect of active vibration suppression is examined in the improvement of image fusion.

To the best of our knowledge, this is the first experimental framework that couples distributed super-resolution imaging with AVC on a flexible-wing platform. The main contributions of this work are summarized below:
\begin{itemize}
    \item A distributed imaging concept is introduced in the context of HAPS technology, demonstrating that commercial lightweight camera systems can meet HAPS payload constraints.

    \item A combined super-resolution and vibration control framework is proposed, highlighting the interdependence between structural dynamics and multi-sensor image fusion quality.

    \item A scaled-down experimental wing demonstrator platform is developed to test and validate the framework under realistic structural deformation conditions.
\end{itemize}

\section{The Demonstrator Platform}

We have designed and built a unique wing demonstrator platform, which captures the structural characteristics of real-life HAPS platforms in terms of flexibility and vibration modes. A properly scaled-down version of the platform allows us to study current challenges in HAPS technology, such as onboard imaging or vibration control, in a simple laboratory environment.
The platform simulates a mission with a distributed payload consisting of two lightweight optical sensors positioned along the wing. As in any Earth observation mission, the goal is to maximize image quality in the presence of atmospheric disturbances, using the available sensors and electronics.
A similar idea of a simplified, flexible test platform was discussed in \cite{Li_VisualServoing_2024}, where vibration suppression also plays an important role. In our case, however, since the visual sensors are placed on the flexible platform, vibration control is crucial to obtain high-quality sensor information.

\begin{figure}[htbp]
    \centering
\includegraphics[width=0.6\textwidth]{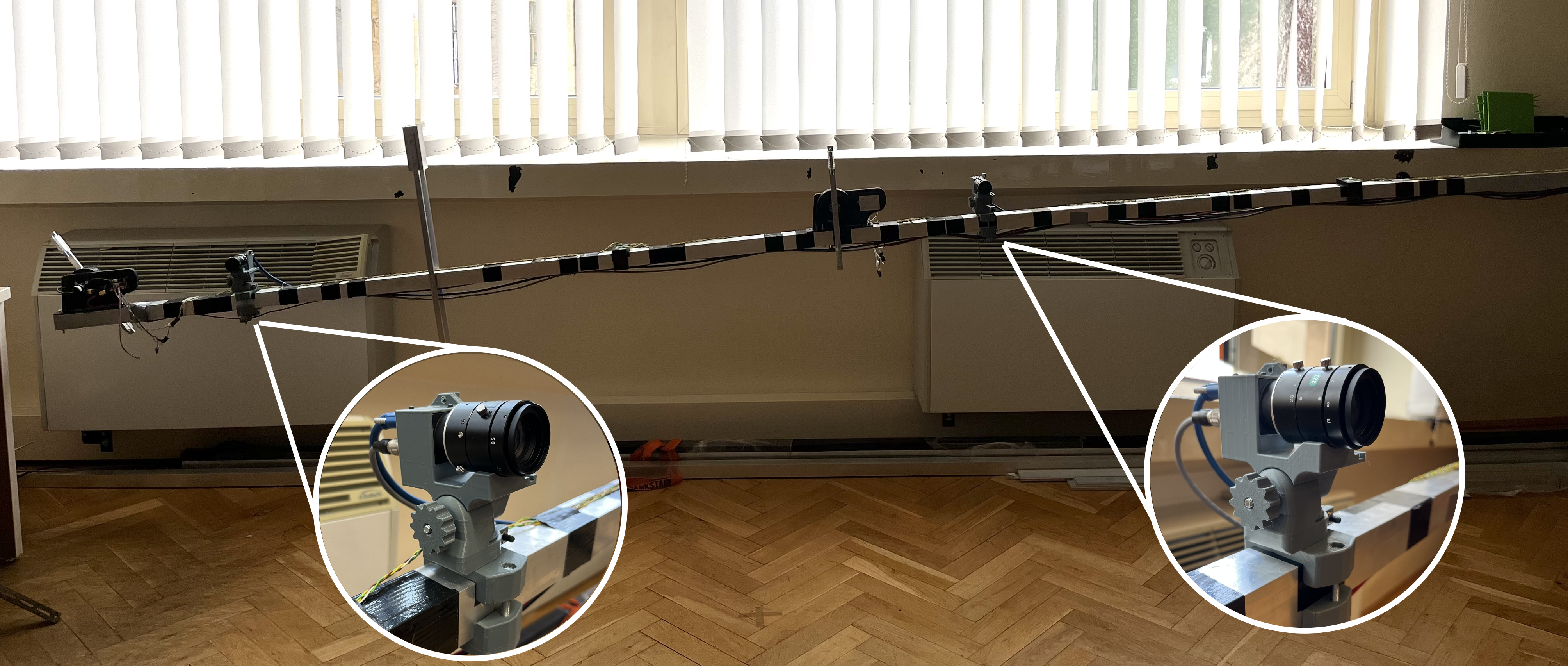}
\caption{The experimental platform equipped with two cameras and actuators.}
\label{fig:hardware_conf}
\end{figure}

\begin{figure}[htbp]
\centering
\includegraphics[width=0.8\textwidth]{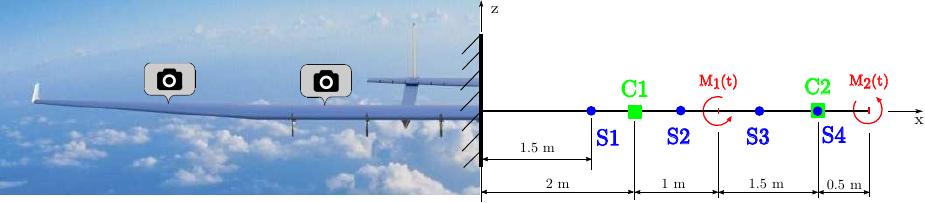}
\caption{Real-life HAPS platform (left), mechanical model with the relevant components (right).} \label{fig:hapsmodel}
\label{fig:setup}
\end{figure}

The platform consists of two actuators, four Inertial Measurement Units (IMUs), and two cameras mounted at different locations. Representing a general case, the cameras are positioned such that all the considered elastic modes of the platform affect their imaging, i.e., neither of them is placed at a vibration node. Fig. \ref{fig:hardware_conf} shows this setup, highlighting the two wing-mounted cameras and DC motors. The corresponding mechanical model is shown in Fig. \ref{fig:setup}.
 Structural vibrations are artificially generated by a gust-emulating actuator, and subsequently suppressed using a secondary active mitigation actuator. The inputs from these motors can be modeled as concentrated moments: $M_1(t)$ represents the unknown disturbance, and $M_2(t)$ handles vibration mitigation based on IMU measurements $S1-S4$. More precisely, the embedded computer receives acceleration and angular velocity data from the IMUs, and sends a control signal to the mitigation actuator. The cameras $C1$ and $C2$ look at predefined patterns placed on the sidewall of the laboratory, representing the observed ground, to provide a controlled and measurable target for image quality assessment.
The recorded images are post-processed with SR techniques, and the resulting image quality is evaluated under various disturbances. Finally, results are compared with those obtained when active vibration mitigation is enabled.

\section{The Proposed Framework}
\subsection{The modeling procedure of the wing platform}

Several design parameters influence the flexible dynamics of the wing platform, including its length, cross-section, and material. While the semi-span of a typical HAPS platform often exceeds 30 meters, our platform's length is limited to 5 meters due to laboratory size constraints. Nonetheless, other design parameters can be adjusted to fit our scaled-down platform.

Assuming a cantilever beam model, the design parameters have been selected so that the low-frequency modes fall within the same range as those of real-world platforms, with special attention given to tuning the first natural frequency. Typically, the first bending mode of a HAPS platform has a frequency below 10 rad/s (1.59 Hz) \cite{shearer:trajectory_cont}.
Based on this criterion, Rayleigh's principle is used to design the cross-section and elastic properties of the platform. Additionally, the designed beam should remain stiff enough to prevent permanent deformations or structural failure. Considering these requirements while minimizing the manufacturing costs, an isotropic, prismatic beam with a hollow-square cross-section is selected, and the material is chosen to be AlMgSi0.5 alloy.

In order to design a proper controller for structural vibration mitigation, the first step is to create a control-oriented model that captures the relevant dynamics of our platform. For this purpose, we carry out finite element (FE) simulations to derive the plant model based on the elastic equation of motion:
\begin{equation} \label{eq:EoM}
\mathbf{\tilde{M}\ddot{q}}+\mathbf{\tilde{C}\dot{q}}+\mathbf{\tilde{K}q}=\mathbf{\tilde{F}},
\end{equation}
where $\mathbf{\tilde{M}}$, $\mathbf{\tilde{C}}$, and $\mathbf{\tilde{K}}$ are the structural matrices transformed into the modal space,
$\mathbf{\tilde{F}}$ is the vector of modal forces, and $\mathbf{q}$ is the vector of modal coordinates. To minimize computational effort, it is common practice to include only the modes within the control bandwidth. The effect of truncated modes can be modeled as uncertainty due to unmodeled dynamics. In this study, the bandwidth is limited to 20 Hz, which corresponds to the first seven modes of the wing, including vertical, lateral, and torsional vibrations.
This relatively broad range was chosen to evaluate the effectiveness of the proposed framework across a variety of disturbance frequencies.

Prior to modal analysis, a nonlinear static analysis was conducted to account for the prestress effects induced by the large deformations due to gravity.
The mathematical model was updated using modal data obtained from experimental vibration tests. During these tests, the system was excited with a sine sweep (0.5–25 Hz), while IMUs recorded accelerations and angular velocities. The identified vibration modes are listed in the left columns of Table \ref{tab:natfreq}. The proximity of lateral and vertical modes can be attributed to the symmetry of the cross-section.

The FE model is optimized based on measurements using the cost function outlined in \cite{sys_id}:
\begin{equation} \label{eq:cf}
\min\limits_{\mathbf{p}} \sum\limits_{i=1}^{n} \left[1-\text{MAC}(\mathbf{\Phi}_{id,i},\mathbf{\Phi}_{fe,i}(\mathbf{p})) \right] + ||\mathbf{e_{\omega}}(\mathbf{p})||,
\end{equation}

which accounts for errors in both the natural frequencies and the mode shape mismatches. The relative error in case of $n$ identified modes is given by
\begin{equation}
\mathbf{e_{\omega}}(\mathbf{p})=\left[ \frac{\omega_{id,1}-\omega_{fe,1}}{\omega_{id,1}} ~~ \ldots ~~
\frac{\omega_{id,n}-\omega_{fe,n}}{\omega_{id,n}}
 \right]^\intercal,
\end{equation}

where $\omega_{id,i}$ and $\omega_{fe,i}$ refer to the $i$\textsuperscript{th} natural frequency of the vibration test and the FE model, respectively.
The Modal Assurance Criterion (MAC) is defined as

\begin{equation}
\text{MAC}(\mathbf{\Phi}_{id,i},\mathbf{\Phi}_{fe,i}(\mathbf{p})) := \frac{|\mathbf{\Phi}_{id,i}^\intercal
\mathbf{\Phi}_{fe,i}|^2}
{\mathbf{\Phi}_{id,i}^\intercal \mathbf{\Phi}_{id,i} \mathbf{\Phi}_{fe,i}^\intercal \mathbf{\Phi}_{fe,i}},
\end{equation}

where $\mathbf{\Phi}_{id,i}$ and $ \mathbf{\Phi}_{fe,i}(\mathbf{p})$ represent the $i$\textsuperscript{th} mode shape vector of the vibration test and the FE model, respectively.
In Eq. (\ref{eq:cf}), the argument $\mathbf{p}$ refers to the tunable parameters of the FE model. Because of the simple geometry and isotropic behavior, the structure can be modeled with high accuracy and therefore, there is no need to tune a large number of uncertain parameters. The optimization process uses a meta-heuristic approach known as differential evolution \cite{diff_ev}. The results are summarized in Table \ref{tab:parameters}.

\begin{table}[htbp]
\caption{Tunable parameters and model optimization results.}
\label{tab:parameters}
\begin{center}
\begin{tabular}{@{}lccc@{}}
\toprule
\textbf{Name} & \textbf{Nominal value} & \textbf{Bounds} & \textbf{Optimized value} \\
\midrule
Equivalent modulus     & 60 GPa     & 55\,--\,65 GPa      & 59 GPa      \\
Poisson's ratio     & 0.3         & 0.25\,--\,0.35      & 0.3         \\
Mass of C1          & 0.55 kg     & 0.5\,--\,0.6 kg     & 0.58 kg     \\
Mass of C2          & 0.55 kg     & 0.5\,--\,0.6 kg     & 0.56 kg     \\
Inertia of actuator & 4.5 gm$^2$  & 1\,--\,10 gm$^2$    & 5.5 gm$^2$  \\
\bottomrule
\end{tabular}
\end{center}
\end{table}

Although the nominal value of the material Young's modulus for the chosen AlMgSi0.5 alloy is $E=70$ GPa, the optimization uses an effective modulus $E_{eq}$ that incorporates unmodelled dynamics, base flexibility, and cabling effects. Static tip-deflection tests showed that the experimental compliance is closely reproduced for $E_{eq}=60$ GPa, which is therefore selected as a nominal value for $E_{eq}$ in the optimization. This choice reflects the actual behavior of the assembled test structure rather than the intrinsic material property.

\begin{figure}[h!]
\centering
\includegraphics[width=0.5\textwidth]{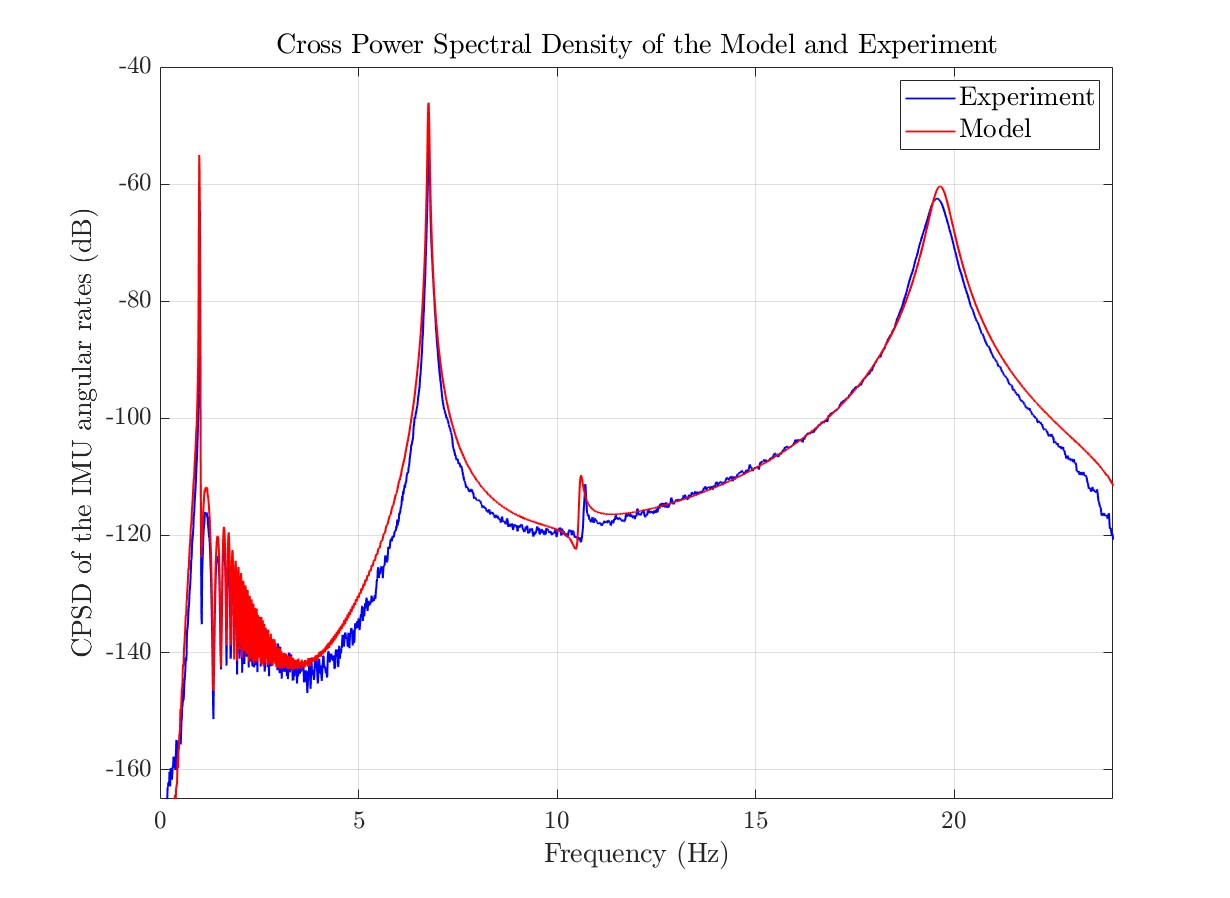}
\caption{CPSD diagram of the structure and the fitted model.}
\label{sys_id}
\end{figure}

The resulting natural frequencies are listed in Table \ref{tab:natfreq}, while the spectrum of the updated finite element model is compared to the experimental spectrum in Fig. \ref{sys_id}.

Proportional damping was assumed in the model, with damping ratios specified in the frequency domain to match the experimental spectrum. The investigated modes are generally highly underdamped; in particular, the first two modes exhibit a damping ratio of 0.09\%.

\begin{table}[h!]
\centering
\caption{Comparison of natural frequencies.}
\label{tab:natfreq}
\renewcommand{\arraystretch}{1.3}
\begin{tabular}{@{}lccc@{}}
\toprule
\textbf{} & \textbf{Experiment (Hz)} & \textbf{Model (Hz)} & \textbf{Rel. Diff. (\%)} \\
\midrule
1\textsuperscript{st} lateral bending     & 0.993 & 0.991 & 0.18 \\
1\textsuperscript{st} vertical bending    & 0.998 & 0.997 & 0.12 \\
2\textsuperscript{nd} lateral bending     & 6.770 & 6.762 & 0.32 \\
2\textsuperscript{nd} vertical bending    & 6.784 & 6.883 & 1.66 \\
1\textsuperscript{st} torsion             & 10.67 & 10.57 & 0.90 \\
3\textsuperscript{rd} lateral bending     & 17.37 & 16.02 & 7.81 \\
3\textsuperscript{rd} vertical bending    & 19.57 & 19.65 & 0.43 \\
\bottomrule
\end{tabular}
\end{table}

In this study, the measurement setup is configured to primarily excite the vertical bending modes. As a result, the relative differences in the lateral bending modes are less significant from a control design point of view. This is also confirmed by the spectral distribution in Fig. \ref{sys_id}, where the vertical bending modes are predominantly amplified on the Cross Power Spectral Density (CPSD) diagram, which is derived from the combined elastic deformation rates registered by the IMUs.

Given the dominant direction of excitation, the model order can be reduced by excluding modes with negligible contribution. Following a conventional modal truncation criterion \cite{ahmad2016modal}, only modes whose cumulative effective modal mass exceeds 90\% are retained. The normalized effective modal mass of the
$i$\textsuperscript{th} mode is defined as
\begin{equation}
    \beta_i=\frac{\gamma_i^2}{m_{total}},
\end{equation}
where $\gamma_i$ is the participation factor of the $i$\textsuperscript{th} mode and $m_{total}$ is the total mass of the structure. For rotational degrees of freedom, $m_{total}$ corresponds to the rotational mass.

As reported in Table \ref{partfact}, the first three vertical bending modes dominate the primary disturbance direction. These modes account for a cumulative $\beta$ of 0.911 for translation and 0.999 for rotation in the vertical plane, indicating that retaining them provides an accurate representation of the system’s response along this direction.

\begin{table}[!htbp]

\caption{Modal participation factors and effective modal masses for model reduction.}
\label{partfact}

\renewcommand{\arraystretch}{1.2}
\centering
\begin{tabular}{lcccc}
\toprule
\textbf{Mode} & $\boldsymbol{\gamma_{tr}}$  & $\boldsymbol{\beta_{tr}}$ & $\boldsymbol{\gamma_{rot}}$  & $\boldsymbol{\beta_{rot}}$\\
\midrule
1\textsuperscript{st} vertical bending & 2.75  & 0.711 & 10.86 & 0.978 \\
2\textsuperscript{nd} vertical bending & -1.29   & 0.157 & -1.50  & 0.019 \\
3\textsuperscript{rd} vertical bending & 0.68  & 0.043 & 0.49  & 0.002 \\
\bottomrule
\multicolumn{2}{r}{\textbf{Sum:}} & 0.911 & \multicolumn{1}{r}{\textbf{Sum:}} & 0.999\\
\end{tabular}
\end{table}

By using this framework, the control synthesis is simplified due to the reduction in model order. The method can be extended to more complex cases, such as those involving mode coupling or multidirectional disturbances.

\subsection{Active vibration control design} \label{ch:AVC}
The designed demonstrator platform offers the opportunity to test different gust load alleviation control strategies, which are commonly used in the aircraft industry, such as PID control, Linear Quadratic Gaussian, or $\mathcal{H}_\infty$ design. While the concept lacks aerodynamics, it provides a structured, scalable framework for studying vibration-related phenomena.
The present paper showcases the control design procedure and related results for the nominal $\mathcal{H}_\infty$ control method.

\begin{figure}[!ht]
\centering
\includegraphics[width=0.4\textwidth]{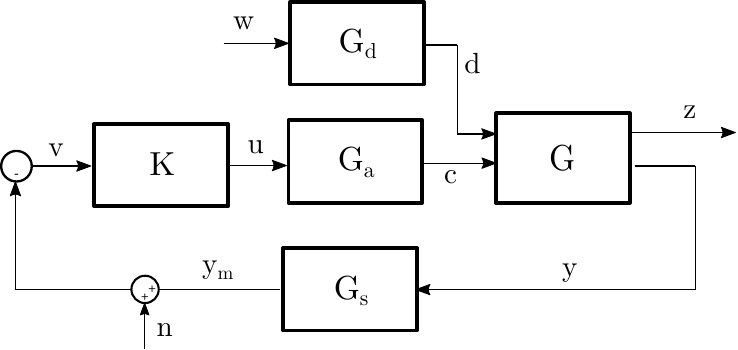}
\caption{The closed-loop architecture.}
\label{state_feedback}
\end{figure}

The control architecture is depicted in Fig. \ref{state_feedback}. $\mathbf{G_{a}}$ and $\mathbf{G_{d}}$ denote the transfer functions of the control and disturbance actuators, respectively, $\mathbf{K}$ is the controller to be designed, and $\mathbf{G_s}$ represents the sensor dynamics. The plant $\mathbf{G}$ has two input channels (disturbance and control moments), forwarding IMU measurements ($\mathbf{y}$) to the controller.

The state space model of $\mathbf{G}$, which is used for control synthesis, has a general structure given by Eq. (\ref{genss}) as
\begin{align} \label{genss}
\mathbf{\dot{x}} & =\mathbf{Ax}+\mathbf{B_1d}+\mathbf{B_2u}, \notag \\
\mathbf{z} & =\mathbf{C_1x}+\mathbf{D_{11}d}+\mathbf{D_{21}u}, \\
\mathbf{y} & = \mathbf{C_2x}+\mathbf{D_{12}d}+\mathbf{D_{22}u}, \notag
\end{align}

where $\mathbf{x}=[\mathbf{q} ~~ \mathbf{\dot{q}}]^\intercal$ denotes the state vector, $\mathbf{w}$ and $\mathbf{u}$ are the disturbance and control inputs, respectively. After the necessary rearrangement of Eq. (\ref{eq:EoM}), the equation of motion can be expressed in state space form as follows:

\begin{align}  \label{eq:SS}
\left[
\begin{array}{c}
\mathbf{\dot{q}} \\
\mathbf{\ddot{q}}
\end{array} \right] &= \underbrace{\left[
\begin{array}{cc}
\mathbf{0} & \mathbf{I} \\
-\mathbf{\tilde{M}}^{-1}\mathbf{\tilde{K}} & -\mathbf{\tilde{M}}^{-1}\mathbf{\tilde{C}}
\end{array} \right]}_\text{$\mathbf{A}$}
\left[
\begin{array}{c}
\mathbf{q} \\
\mathbf{\dot{q}}
\end{array} \right] \notag \\
&+ \underbrace{\left[
\begin{array}{c}
\mathbf{0} \\
\mathbf{\tilde{M}^{-1}}
\end{array} \right]}_\text{$\mathbf{B_1}$}
\begin{array}{c}
 \left[
\mathbf{\mathbf{\tilde{F}_d}}
\right]  \\
\end{array}+ \underbrace{\left[
\begin{array}{c}
\mathbf{0} \\
\mathbf{\tilde{M}^{-1}}
\end{array} \right]}_\text{$\mathbf{B_2}$}
\begin{array}{c}
 \left[
\mathbf{\mathbf{\tilde{F}_u}}
\right] \\
\end{array},
\end{align}

where $\mathbf{\tilde{M}}, \mathbf{\tilde{K}}, \mathbf{\tilde{C}} \in \mathbb{R}^{3 \times 3}$, $\mathbf{\tilde{F}} \in \mathbb{R}^{3 \times 1}$, and due to the state reduction, $\mathbf{q} \in \mathbb{R}^{3\times1}$. Note that the separation $\mathbf{\tilde{F}}=\mathbf{\tilde{F}_d}+\mathbf{\tilde{F}_u}$ is performed because of the distinct disturbance and control actions.

In addition to preserving structural integrity, the primary goal of the control design is to minimize image degradation. Therefore, the motion blur caused by structural vibrations needs to be minimized, which is best described by the elastic deformation rates ($\mathbf{\dot{d}_e}$) at the locations of the cameras.  For this reason, the performance output vector ($\mathbf{z}$) in Fig. \ref{state_feedback} is defined by these deformation rates as
\begin{equation}
\mathbf{z}=\mathbf{\dot{d}_{e,cam}}=\left[
\begin{array}{cc}
\dot{d}_{e,C1} & \dot{d}_{e,C2} \\
\end{array} \right] ^\intercal .
\end{equation}

  Furthermore, the output vector $\mathbf{y}$ is constructed such that it contains angular rate ($\mathbf{\dot{d}_{e,imu}}$) and linear acceleration measurements ($\mathbf{a_{imu}}$) from the available IMUs.

The characteristics of the onboard actuators $G_d$ and $G_a$ were identified by experiments from command current signal to output torque, with a two-pole discrete transfer function fitted to each dataset.

The IMU measurements are affected by noise, which is modeled as white noise drawn from Gaussian distribution and characterized by $\mathbf{n} \sim \mathcal{N}(\mathbf{0},\mathbf{R}_n)$.
The covariance matrix of the sensor noise is determined by static noise measurements. Sensor dynamics $G_s$ are modeled as a first-order high-pass filter with cut-off frequency at 0.1 Hz for each channel.
Based on previous experiences with project FliPASED \cite{flipased}, the sampling frequency was set to 300 Hz to balance computational cost against control performance degradation (e.g., due to discretization errors and phase lag). At this sampling rate, preliminary real-time simulations revealed a two-sample delay in the closed-loop, which is therefore incorporated into the model.

The controller receives angular rate measurements from $S1$ and $S3$, and linear accelerations from $S2$ and $S4$, according to Fig. \ref{fig:setup}, and forwards the current command signal to the control actuator motor.
Control action is constrained to avoid actuator saturation, introducing a trade-off between performance and hardware limitations.

\begin{figure}[h!]
  \centering
  \includegraphics[width=1\linewidth]{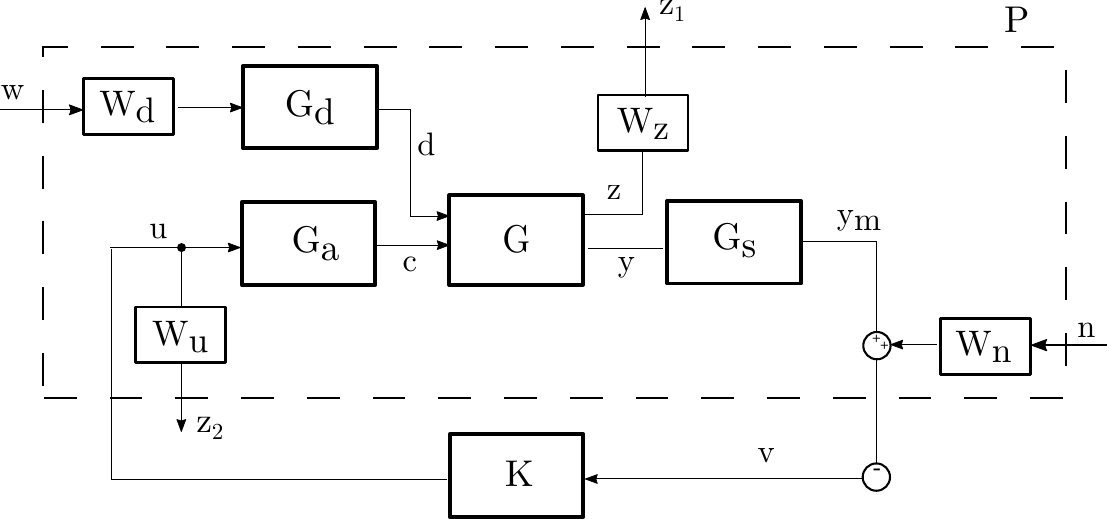}
  \caption{The control architecture extended by normalizing weight functions.}
  \label{fig:genplant}
\end{figure}

 In order for the $\mathcal{H}_\infty$ control design to satisfy the performance specifications, frequency-dependent weight functions are introduced such that the elements of the resulting transfer function matrix have comparable magnitudes inside the control bandwidth, which is specified between 0.5 and 23 Hz.
 The augmented plant structure is shown in Fig. \ref{fig:genplant}.

The initial weighting functions were designed based on simple expressions with clear physical interpretation i.e., scaling the signals to their expected order of magnitude. The weighting parameters were subsequently fine-tuned empirically to achieve improved control performance.
\begin{itemize}
\item
The disturbance weight function is based on the maximum allowable driving current limit of the utilized BLDC motor, which is 60 A:
\begin{equation}
    W_d=60 ~\text{A}.
\end{equation}
\item
The control action is weighted at the output by a band-stop filter, which reduces the penalty in the specified frequency band while limiting excessive actuation at very low and high frequencies. It is implemented as a cascade of first-order filters as:
\begin{equation} \label{eq:Wu}
    W_u=\frac{1}{2}\left(1-\left(\frac{2\pi}{s+2\pi} \right) \left(\frac{s}{s+5\cdot 2\pi}\right)\right) ~\text{A}^{-1},
\end{equation}
which is then discretized using Tustin's method.
\item The noise inputs are weighted by the covariance of the sensor noise obtained from static noise tests and datasheet values:
\begin{equation}
\mathbf{W_n}=10^{-6.3}\mathbf{I_4} ~\text{SI},
\end{equation}
where $\mathbf{I_4}$ denotes the $4\times4$ identity matrix.

\item Preliminary open-loop simulations showed that elastic rates at the camera locations range between 0.01 and 0.1 rad/s. Initially, the performance weights were normalized by the inverse of these rates, and subsequently fine-tuned for enhanced control performance as:
\begin{equation}
    \mathbf{W_z}=\text{diag}\left(10^{1.3} \quad 10^{2.1} \right) ~ \mathrm{s\,rad^{-1}}.
\end{equation}
\end{itemize}

As a next step, let us define a simplified notation for the different transfers in the augmented plant in Fig. \ref{fig:genplant}:
\[
\begin{matrix}
\mathbf{\hat{G}_{11}} \triangleq \mathbf{G_{d \rightarrow z}G_d}
& \quad &
\mathbf{\hat{G}_{12}} \triangleq \mathbf{G_s G_{d \rightarrow y}G_d}
\\[1ex]
\mathbf{\hat{G}_{21}} \triangleq \mathbf{G_{c \rightarrow z}G_a}
& \quad &
\mathbf{\hat{G}_{22}} \triangleq \mathbf{G_s G_{c \rightarrow y}G_a}
\end{matrix}
\]
where the arrows in subscripts denote the corresponding plant transfers.

The augmented open-loop system interconnection matrix $\mathbf{P}$ incorporates the transfers from all inputs to outputs in accordance with Fig. \ref{fig:genplant}. With appropriate partitioning, this can be written as
\begin{equation}
         \left[
     \begin{array}{c}
          \mathbf{z_1} \\ \mathbf{z_2} \\ \mathbf{v} \\
     \end{array} \right]=
     \left[
     \begin{array}{c|c}
          \mathbf{P_{11}}  & \mathbf{ P_{12}} \\ \hline
         \mathbf{ P_{21}} & \mathbf{P_{22}} \\
     \end{array} \right]
     \left[
     \begin{array}{c}
         \mathbf{w} \\ \mathbf{n} \\  \mathbf{u} \\
     \end{array} \right].
\end{equation}
With the introduced notations and weight functions listed above, $\mathbf{P}$ becomes:

\begin{equation} \label{eq:genplant}
     \mathbf{P}=\text{diag}\left(
     \begin{array}{c}
          \mathbf{W_z} \\ \mathbf{W_u} \\
          \hline
          \mathbf{I} \\
     \end{array} \right)
     \left[
     \begin{array}{cc | c}
          \mathbf{\hat{G}_{11}} & \mathbf{0} & \mathbf{ \hat{G}_{21}} \\
         \mathbf{0} & \mathbf{0} & \mathbf{I} \\
         \hline
         -\mathbf{\hat{G}_{12}} & \mathbf{-I} & -\mathbf{\hat{G}_{22}} \\
     \end{array} \right]
     \text{diag} \left(
     \begin{array}{c}
          \mathbf{W_d} \\ \mathbf{W_n} \\ \hline \mathbf{I} \\
     \end{array} \right)
 \end{equation}

 Partitioning $\mathbf{P}$ in Eq. (\ref{eq:genplant}) according to the drawn lines, the lower linear fractional transformation $\mathcal{F}_l(\mathbf{P},\mathbf{K})$ determining the required closed-loop transfer functions can be expressed as
 \begin{equation} \label{eq:lft}
\mathcal{F}_l(\mathbf{P},\mathbf{K})=\mathbf{P_{11}}+\mathbf{P_{12}K}(\mathbf{I}-\mathbf{P_{22}K})^{-1}\mathbf{P_{21}}.
 \end{equation}

 The control goal is to synthesize a nominal controller $\mathbf{K}$ such that the induced $\mathcal{L}_{2}$-norm of the derived closed-loop transfer function matrix is minimized, i.e.,
 \begin{equation} \label{eq:gammaopt}
     \min_\mathbf{K} ~\gamma \quad \quad \text{s.t.} \quad \quad ||\mathcal{F}_l(\mathbf{P},\mathbf{K})||_{\mathcal{L}_{2,i}} < \gamma.
 \end{equation}

\section{Distributed-Camera Super-Resolution Methodology}
\label{sec:super-resolution}

\begin{figure}[htbp]
  \centering
  \includegraphics[width=0.95\linewidth]{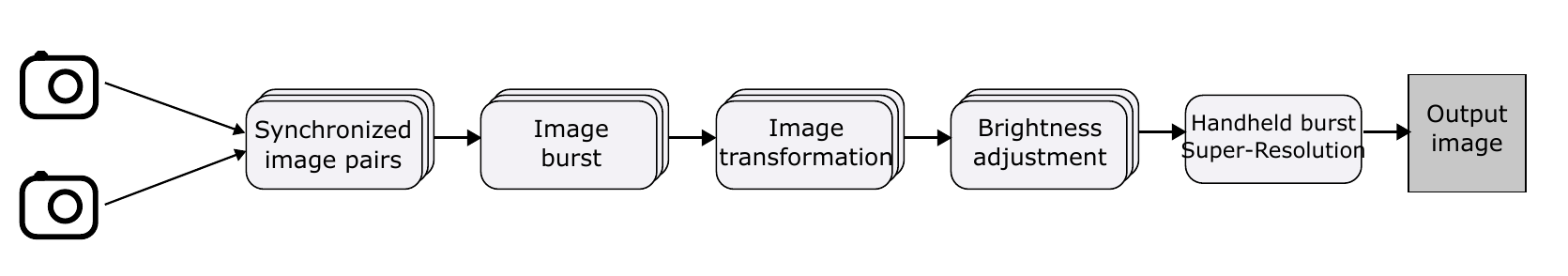}
  \caption{The proposed super-resolution pipeline. The synchronized image pairs are divided into image bursts. After image transformation and brightness adjustment, the burst is fed into the handheld super-resolution algorithm introduced in \cite{lafenetre2023implementing}.}
  \label{fig:sr-pipeline}
\end{figure}

We have developed a pipeline for a super-resolution (SR) algorithm tailored for processing images from the multi-camera system. The proposed image-processing pipeline extends the SR algorithm introduced in \cite{lafenetre2023implementing}, originally developed for handheld cameras such as mobile phones, by adapting it to effectively handle images captured by the multi-camera setup. The complete process is illustrated in Fig.~\ref{fig:sr-pipeline}. The pipeline starts by collecting synchronized image pairs, which are then passed through the preprocessing steps detailed below. Adding these preprocessing steps enables the handheld burst method to be applied to a distributed camera system operating at high altitude.

\begin{enumerate}
\item \textbf{Synchronized image pairs.}
Synchronizing image acquisition across cameras ensures that only frames captured within a narrow temporal window are grouped into a burst, thereby preserving spatial correspondence between views. This preprocessing step is necessary because images captured at substantially different times may cover different areas and therefore cannot be meaningfully fused within a burst.

\item \textbf{Image burst.}
The test images are divided into bursts of a fixed size. Sample bursts from the recorded image set were processed, and SR quality was evaluated with the nine metrics listed in Sec.~\ref{sec:evaluation-metrics}. For each progressively increasing burst size, scores were averaged over five selected sequences. Quality improved steadily up to 16 images per camera, beyond which the additional improvement was not relevant. Eight of the nine metrics show no noticeable improvement beyond this threshold, while smaller bursts degrade quality. DISTS behaves differently, reaching its optimum around eight images per camera, but its subsequent quality drop is marginal relative to the gains across the other eight metrics. Thus, 16 images per camera provided a good compromise between output quality and computational cost and was used in all subsequent quality evaluations. Figure~\ref{fig:burst-size} presents the complete nine-metric test.

\begin{figure}[htbp]
  \centering
  \begin{subfigure}[t]{0.31\linewidth}
    \includegraphics[width=\linewidth]{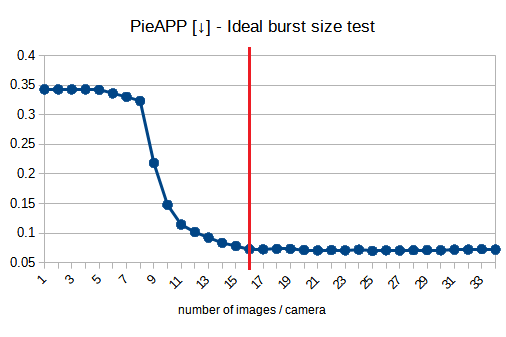}
    \caption{PieAPP}
  \end{subfigure}\hfill
  \begin{subfigure}[t]{0.31\linewidth}
    \includegraphics[width=\linewidth]{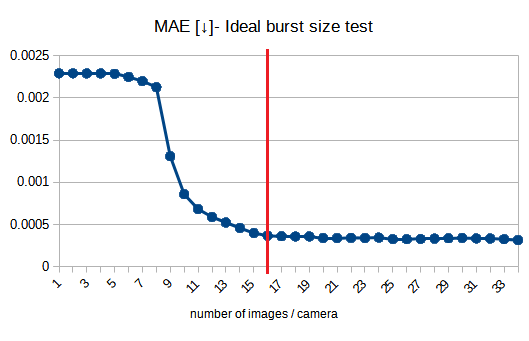}
    \caption{MAE}
  \end{subfigure}\hfill
  \begin{subfigure}[t]{0.31\linewidth}
    \includegraphics[width=\linewidth]{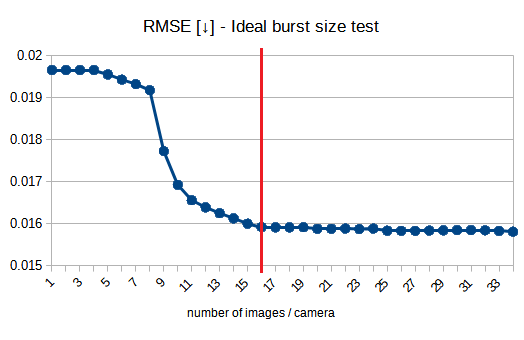}
    \caption{RMSE}
  \end{subfigure}

  \begin{subfigure}[t]{0.31\linewidth}
    \includegraphics[width=\linewidth]{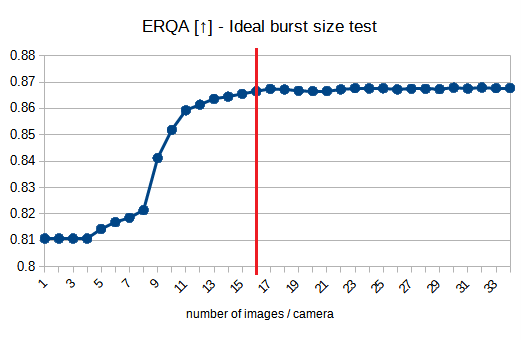}
    \caption{ERQA}
  \end{subfigure}\hfill
  \begin{subfigure}[t]{0.31\linewidth}
    \includegraphics[width=\linewidth]{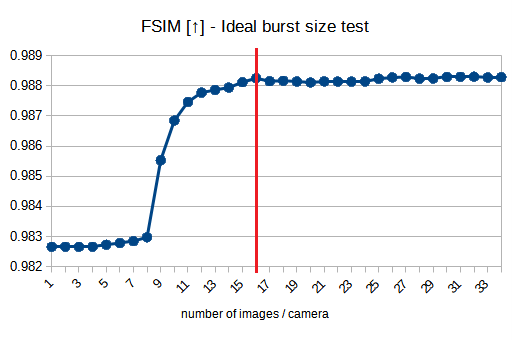}
    \caption{FSIM}
  \end{subfigure}\hfill
  \begin{subfigure}[t]{0.31\linewidth}
    \includegraphics[width=\linewidth]{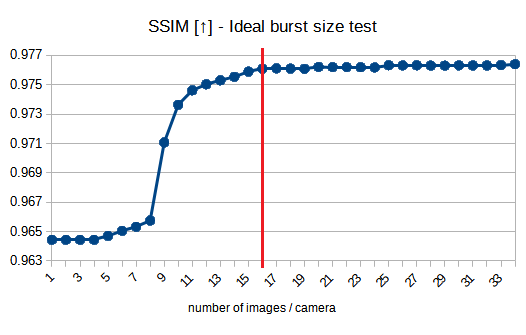}
    \caption{SSIM}
  \end{subfigure}

  \begin{subfigure}[t]{0.31\linewidth}
    \includegraphics[width=\linewidth]{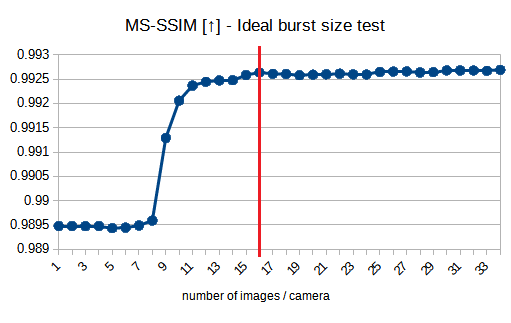}
    \caption{MS-SSIM}
  \end{subfigure}\hfill
  \begin{subfigure}[t]{0.31\linewidth}
    \includegraphics[width=\linewidth]{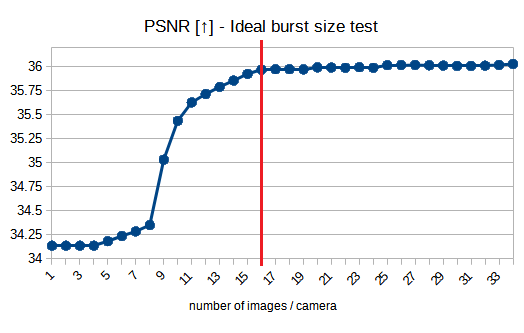}
    \caption{PSNR}
  \end{subfigure}\hfill
  \begin{subfigure}[t]{0.31\linewidth}
    \includegraphics[width=\linewidth]{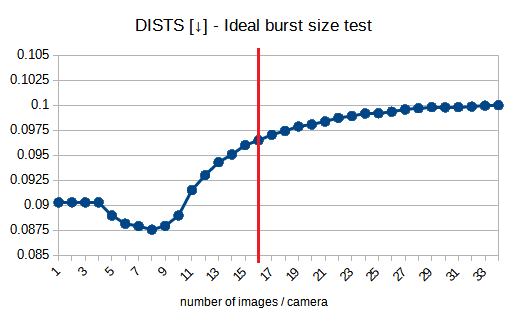}
    \caption{DISTS}
  \end{subfigure}
  \caption{Effect of burst size on the nine evaluation metrics. The selected value of 16 images per camera is marked by a red line. Arrows indicate whether lower ($\downarrow$) or higher ($\uparrow$) values correspond to better results.}
  \label{fig:burst-size}
\end{figure}

\item \textbf{Image transformation.}
The first image from the primary camera serves as the reference, and the remaining burst images are aligned to it. Once SIFT features are extracted, the translation between images is computed and rounded to the nearest integer value before the resulting shift is applied. This step aligns images captured from different viewpoints to a common reference; without it, the differences are too large for the SR algorithm to combine the laboratory images effectively. In real-world scenarios, the greater distance between viewpoints and the target area reduces these differences, making burst processing feasible, although accurately replicating or quantifying that real-world effect in the laboratory remains challenging.

\item \textbf{Brightness adjustment.}
The intensity of each image in the burst is modified so that its average intensity equals that of the reference image, balancing lighting changes. This normalization is particularly important for metric-based evaluations because several image-quality metrics are sensitive to illumination changes. Figures~\ref{fig:brightness-images} and~\ref{fig:brightness-metric} compare processing with and without this step. Omitting the adjustment yields noticeably worse metric values despite no perceptible difference in visual quality.

\begin{figure}[htbp]
  \centering
  \begin{subfigure}[t]{0.48\linewidth}
    \includegraphics[width=\linewidth]{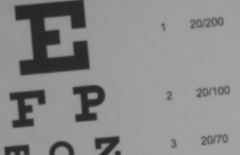}
    \caption{GT without adjustment.}
  \end{subfigure}\hfill
  \begin{subfigure}[t]{0.48\linewidth}
    \includegraphics[width=\linewidth]{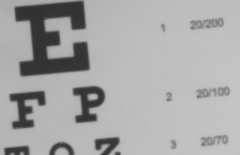}
    \caption{SR output without adjustment.}
  \end{subfigure}

  \begin{subfigure}[t]{0.48\linewidth}
    \includegraphics[width=\linewidth]{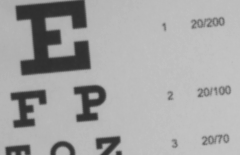}
    \caption{GT with adjustment.}
  \end{subfigure}\hfill
  \begin{subfigure}[t]{0.48\linewidth}
    \includegraphics[width=\linewidth]{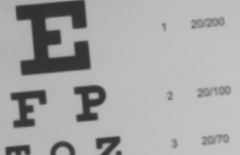}
    \caption{SR output with adjustment.}
  \end{subfigure}
  \caption{Brightness adjustment of the ground-truth (GT) and super-resolved (SR) images. Without adjustment, the GT and SR brightness levels differ significantly; adjustment makes their intensities comparable.}
  \label{fig:brightness-images}
\end{figure}

\begin{figure}[htbp]
  \centering
  \includegraphics[width=0.75\linewidth]{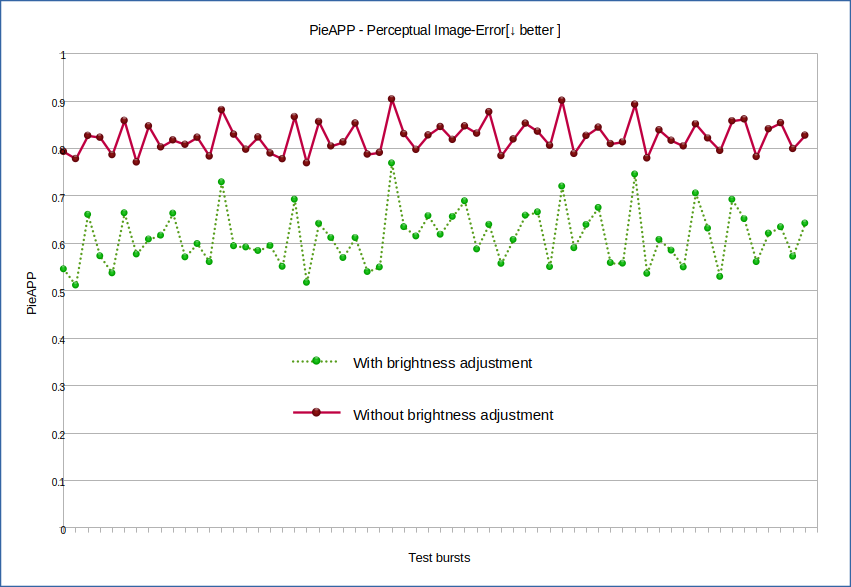}
  \caption{Effect of brightness adjustment on PieAPP quality for different bursts. Omitting brightness adjustment while retaining the rest of the pipeline significantly degrades the metric scores.}
  \label{fig:brightness-metric}
\end{figure}

Across the 62 test bursts in the ``with AVC'' scenario, brightness adjustment clearly improves MAE, RMSE, PSNR, SSIM, MS-SSIM, FSIM, and PieAPP. ERQA and DISTS degrade slightly, but the changes fall within the measurement standard deviations. Table~\ref{tab:brightness-adjustment} reports all nine results.

\begin{table}[htbp]
\centering
\caption{Impact of brightness adjustment on image-quality metrics. Arrows indicate whether higher ($\uparrow$) or lower ($\downarrow$) values are preferred. Results are mean $\pm$ standard deviation.}
\label{tab:brightness-adjustment}
\small
\begin{tabular}{lcc}
\toprule
\textbf{Metric} & \textbf{Without adjustment} & \textbf{With adjustment} \\
\midrule
MAE [$\downarrow$]     & 0.110 $\pm$ 8.20e-4  & \textbf{0.011 $\pm$ 5.89e-4} \\
RMSE [$\downarrow$]    & 0.118 $\pm$ 1.92e-3  & \textbf{0.028 $\pm$ 7.69e-3} \\
PSNR [$\uparrow$]      & 18.549 $\pm$ 1.41e-1 & \textbf{31.5 $\pm$ 2.60} \\
SSIM [$\uparrow$]      & 0.958 $\pm$ 2.22e-3  & \textbf{0.972 $\pm$ 3.12e-3} \\
ERQA [$\uparrow$]      & \textbf{0.860 $\pm$ 1.66e-2} & 0.842 $\pm$ 1.23e-2 \\
MS-SSIM [$\uparrow$]   & 0.978 $\pm$ 4.57e-4  & \textbf{0.990 $\pm$ 7.63e-4} \\
DISTS [$\downarrow$]   & \textbf{0.082 $\pm$ 1.17e-2} & 0.087 $\pm$ 1.38e-2 \\
FSIM [$\uparrow$]      & 0.979 $\pm$ 1.12e-3  & \textbf{0.990 $\pm$ 1.15e-3} \\
PieAPP [$\downarrow$]  & 0.823 $\pm$ 3.26e-2  & \textbf{0.612 $\pm$ 5.88e-2} \\
\bottomrule
\end{tabular}
\end{table}
\end{enumerate}

After preprocessing, the image burst is fed into the handheld burst SR algorithm \cite{lafenetre2023implementing}; the preprocessing allows that algorithm to handle images from two cameras. Because alignment is independent of sensor count, the pipeline can be extended to additional cameras by transforming each viewpoint to a reference image within each burst. The optimal burst size for a sensor configuration can be determined in advance without affecting real-time operation. Processing time and memory use increase linearly with the number of input images; however, currently available embedded computers can meet all performance requirements. For example, the Jetson AGX Orin can support up to six directly connected cameras and has the memory and computational capacity to process the captured images. In practice, the payload capacity of lightweight HAPS platforms, rather than available computational resources, typically limits the camera count.

\section{Experimental Protocol}
\label{sec:experimental-protocol}

The experiment has two distinct input paths. The \emph{hardware/control inputs} are the commanded currents sent to the disturbance and vibration-mitigation actuators and the inertial measurements returned to the controller. The \emph{image-processing inputs} are synchronized frames from the two cameras; these frames are grouped, aligned, brightness-normalized, and passed to the SR algorithm as described in Sec.~\ref{sec:super-resolution}. Keeping these paths distinct makes it possible to relate vibration-control conditions to image quality without treating actuator commands as SR inputs.

\subsection{Control excitation and actuator operating envelope}

\begin{figure}[htbp]
  \centering
  \begin{subfigure}[t]{0.48\linewidth}
    \includegraphics[width=\linewidth]{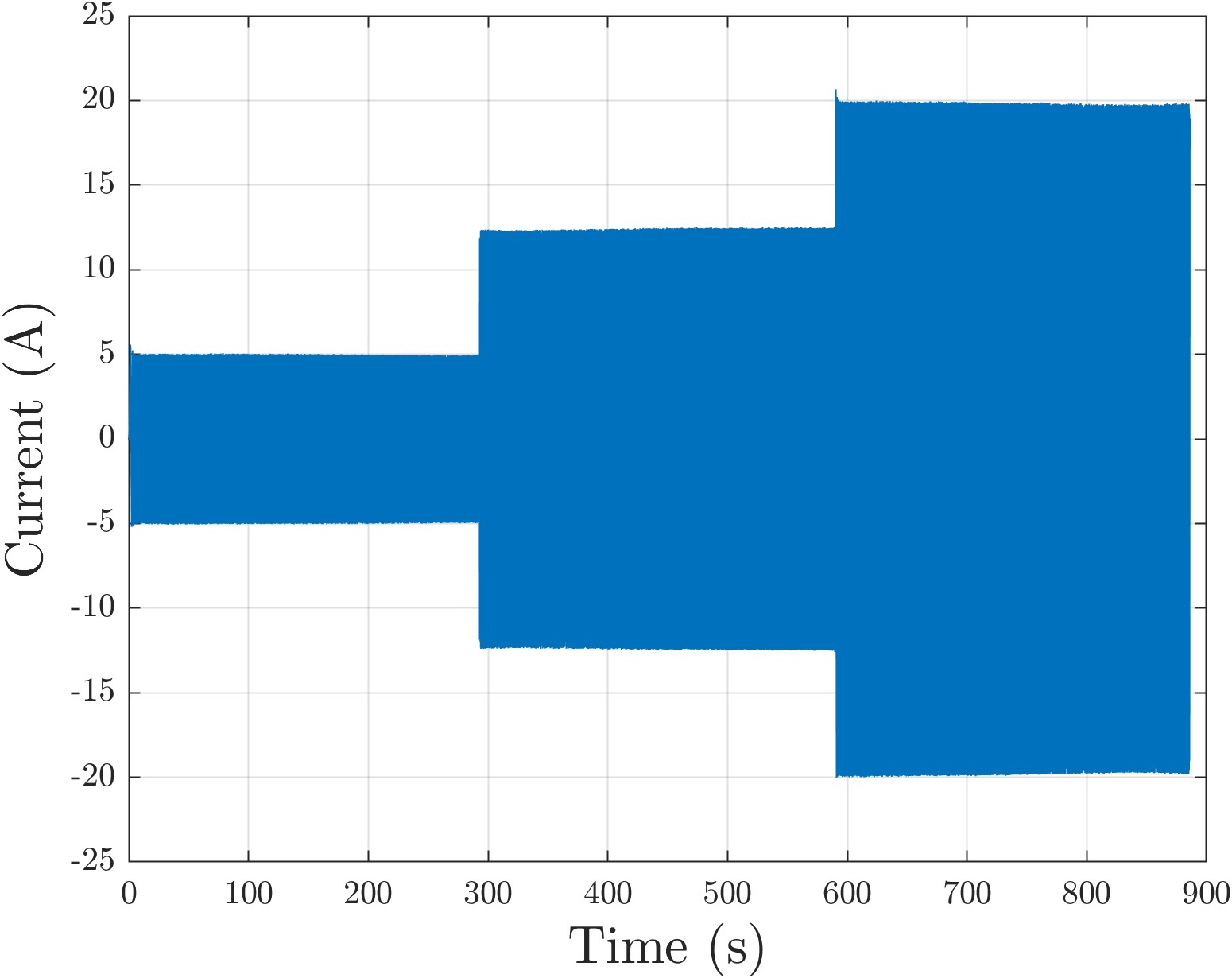}
    \caption{Sine-sweep disturbance input.}
    \label{fig:actuator-disturbance-sweep}
  \end{subfigure}\hfill
  \begin{subfigure}[t]{0.48\linewidth}
    \includegraphics[width=\linewidth]{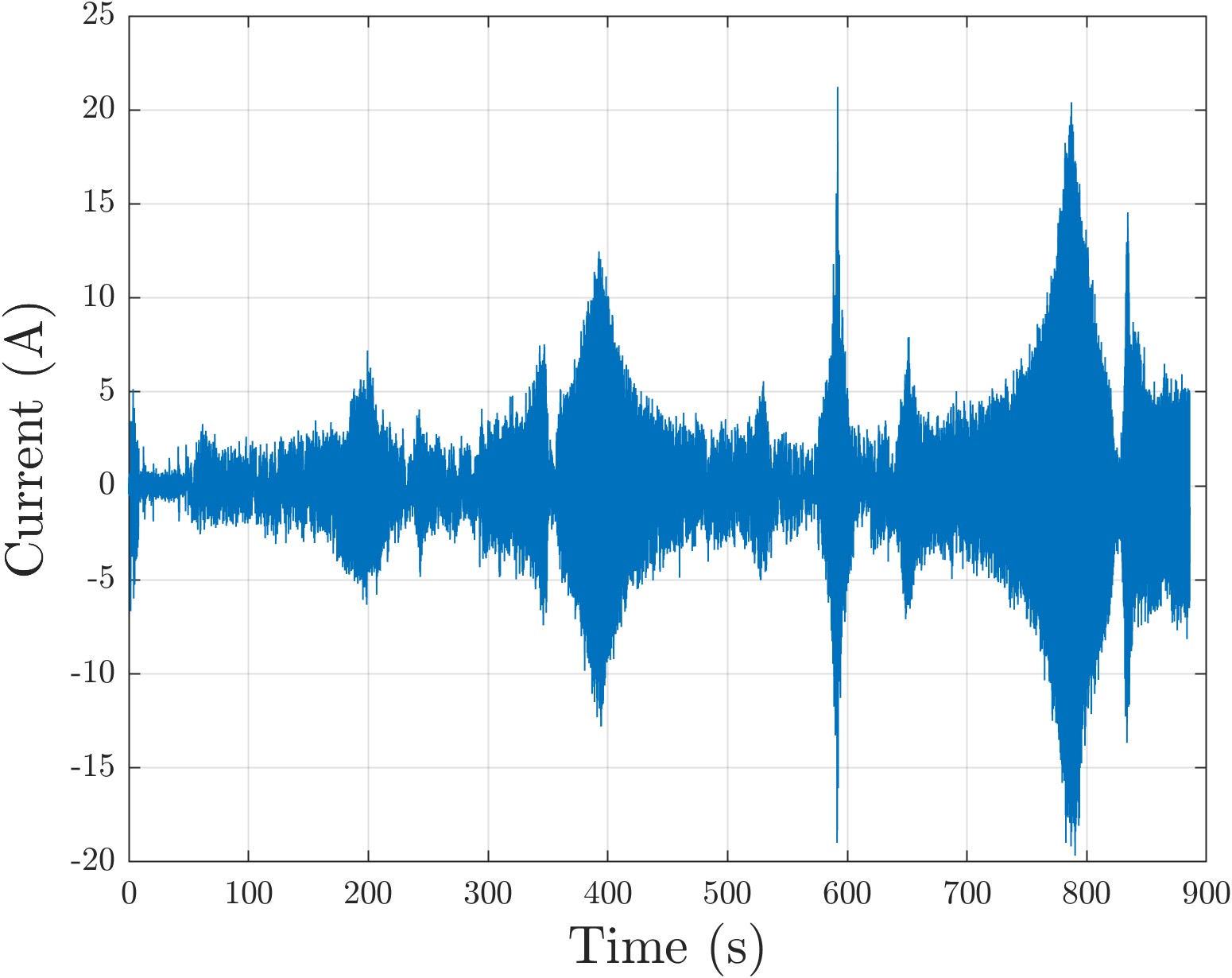}
    \caption{Control command for vibration mitigation.}
    \label{fig:actuator-control-command}
  \end{subfigure}
  \caption{Time-domain actuator signals during the chirp tests.}
  \label{fig:actuator-signals}
\end{figure}

\begin{figure}[htbp]
  \centering
  \begin{subfigure}[t]{0.48\linewidth}
    \includegraphics[width=\linewidth]{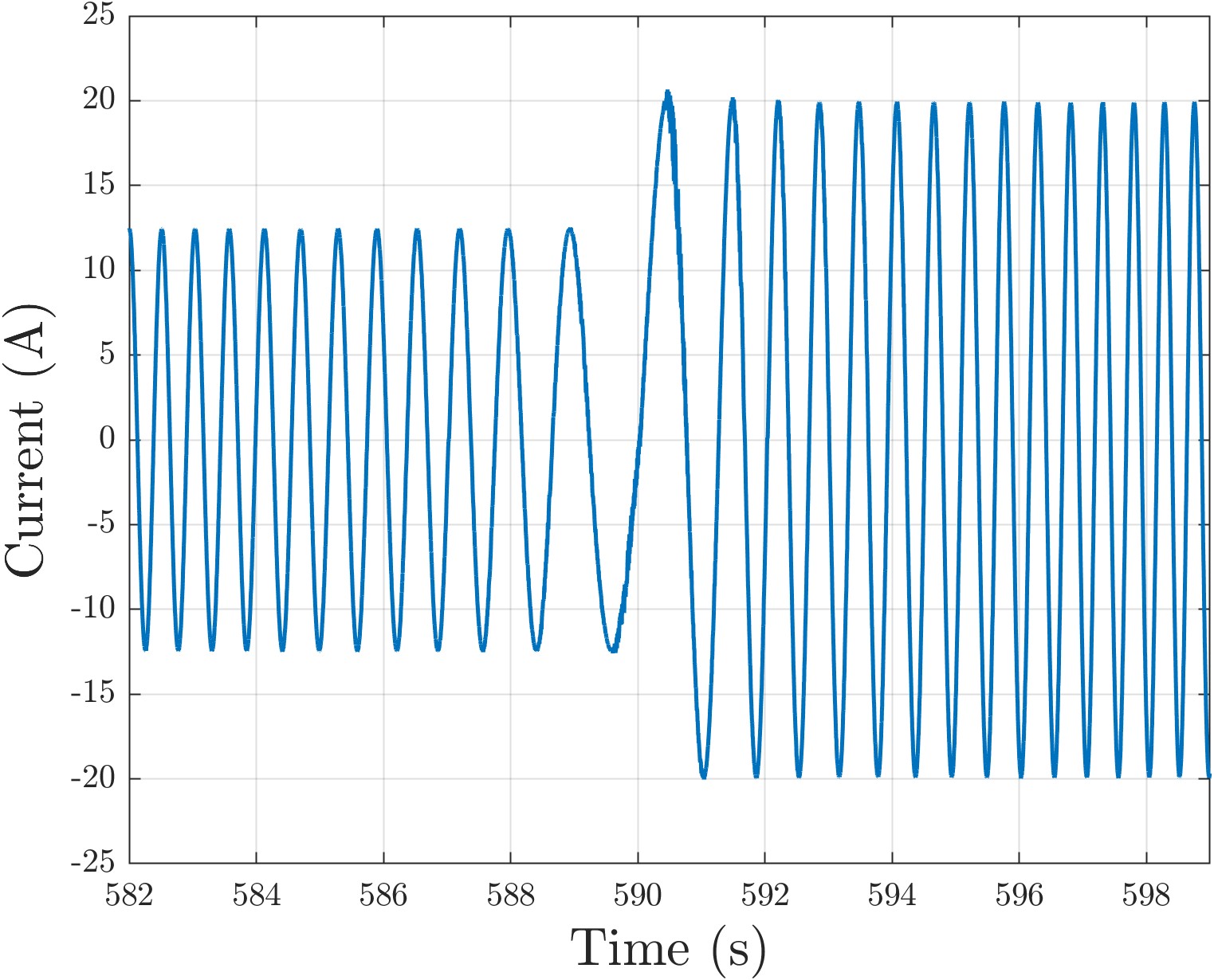}
    \caption{Input sine-sweep disturbance.}
  \end{subfigure}\hfill
  \begin{subfigure}[t]{0.48\linewidth}
    \includegraphics[width=\linewidth]{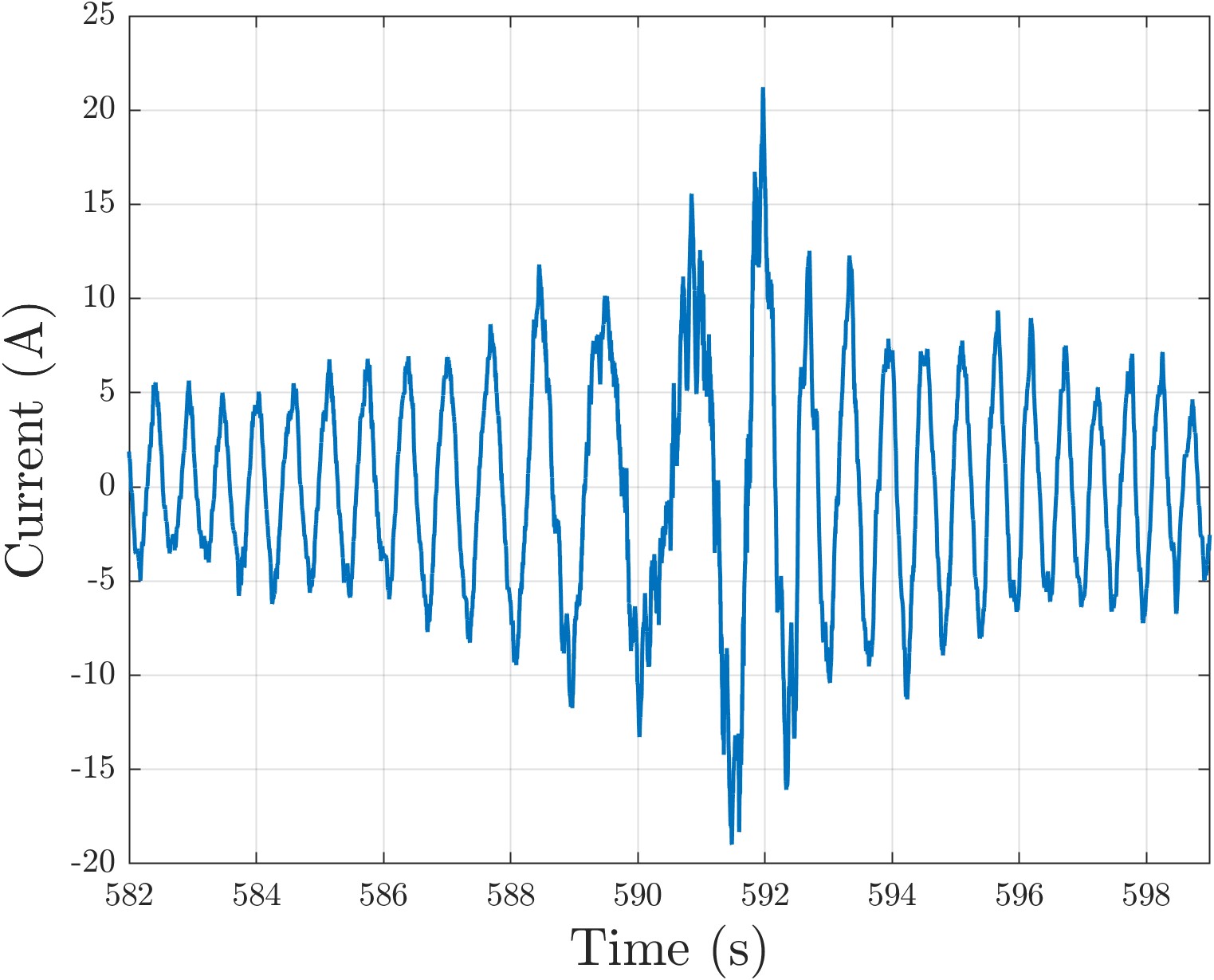}
    \caption{Control-command current.}
  \end{subfigure}
  \caption{Zoomed view of the experimental actuator signals in Fig.~\ref{fig:actuator-signals} during a chirp-magnitude change.}
  \label{fig:actuator-detail}
\end{figure}

Control performance is evaluated across all frequencies within the control bandwidth using a sufficiently long chirp excitation. The disturbance is a sine sweep from 0.5 to 25~Hz with three amplitudes, as shown in Fig.~\ref{fig:actuator-signals}; each chirp lasts $300$~s to provide sufficient frequency resolution. The corresponding actuator command mitigates vibrations induced by the disturbance actuator. The zoomed interval in Fig.~\ref{fig:actuator-detail} exposes the actuator behavior during a chirp-magnitude change. This chirp test provides the baseline for identifying frequency-dependent control performance. Disturbance signals representing atmospheric conditions more directly, such as wind gusts or turbulence, can be considered in further experiments.

The motor hardware saturates at $60$~A. Under the severe disturbance profiles, active-control demand peaks at approximately $21$~A, leaving a $65\%$ operational safety margin before saturation. Because nominal experimental operation remains within this region, the initial prototype omits explicit anti-windup. Under generic atmospheric conditions the threshold may be exceeded; integrating a standard observer-based or back-calculation anti-windup scheme into the $\mathcal{H}_\infty$ framework therefore remains a priority for future work.

\subsection{Image-processing inputs and datasets}

For every scenario, synchronized acquisition ensures that paired frames come from a narrow temporal window. The pairs are formed into bursts of 16 images per camera. The first primary-camera image defines the reference; SIFT-derived translations align the other images to it after integer rounding, and mean intensity is normalized to the reference before SR. These are image-processing operations, whereas the disturbance excitation and mitigation current remain hardware/control inputs.

Structural vibrations have a significant impact on the onboard imaging capability. Control performance is first evaluated at frequencies inside the control bandwidth using inertial sensors. The SR algorithm is then assessed at the disturbance frequency with the highest spectral density, representing a worst-case scenario. According to Fig.~\ref{sys_id}, this frequency is $6.78$~Hz, so it is applied to the disturbance actuator in the imaging tests. The effect of AVC on image quality in the two-camera setup is evaluated under three dataset scenarios:

\begin{enumerate}[label=(\arabic*)]
\item \textbf{Disturbance-free:} the wing is stationary in its reference position and is not subjected to external disturbances.
\item \textbf{Without AVC:} structural excitation is active, but the designed controller is deactivated.
\item \textbf{With AVC:} the designed AVC mitigates vibrations caused by the disturbance.
\end{enumerate}

\subsection{Region of interest, ground truth, and quantitative protocol}

Because the platform has no high-resolution reference sensor apart from the two cameras used by the SR pipeline, a directly captured high-resolution reference is unavailable. For laboratory quantitative evaluation, a $1024\times1024$ window is therefore extracted from each burst image as the region of interest. A window is defined separately for each camera. The first-camera region is the reference; the corresponding second-camera region is determined with the transformation calculated from the first image pair in the dataset, using the same procedure as the transformation step in Sec.~\ref{sec:super-resolution}.

The $1024\times1024$ ground-truth window is also aligned to the reference and brightness-adjusted for comparability. The burst inputs are resized to $512\times512$, and the SR pipeline produces a $1024\times1024$ output for comparison with the aligned ground-truth window. Window selection is needed only because laboratory images may contain objects at different depths and distorted edges, while translation requires a planar region for acceptable performance. At high operating altitude this selection can be removed. Likewise, downsampling is used only to evaluate the SR algorithm and is not part of its normal workflow.

To establish a reliable ground truth under variable lighting and environmental conditions, multiple frames are captured with a fixed camera in the disturbance-free scenario. No single frame contains unique information, so corresponding pixels are averaged. This mimics long exposure, reduces random noise, and produces the ground-truth image used for the tests.

\subsection{Evaluation metrics}
\label{sec:evaluation-metrics}

Each scenario is evaluated with nine frequently applied metrics. MAE, RMSE, PSNR, and SSIM are included as established measures \cite{CompReview_SuperRes_metrics_2024,DL-based_SR_building_height_2024,ImgSR_Reconstr_Improved_GAN_2019,ImgQualityMetrics_PSNR_SSIM_2010}. The more recently introduced DISTS \cite{dists}, ERQA \cite{erqa}, FSIM \cite{fsim}, MS-SSIM \cite{ms_ssim}, and PieAPP \cite{pieapp} complement them. Together, these measures cover pixel accuracy, edge quality, structural and texture preservation, and perceptual quality.

\subsection{Visual-test protocol}

\begin{figure}[htbp]
  \centering
  \begin{subfigure}[t]{0.48\linewidth}
    \includegraphics[width=\linewidth]{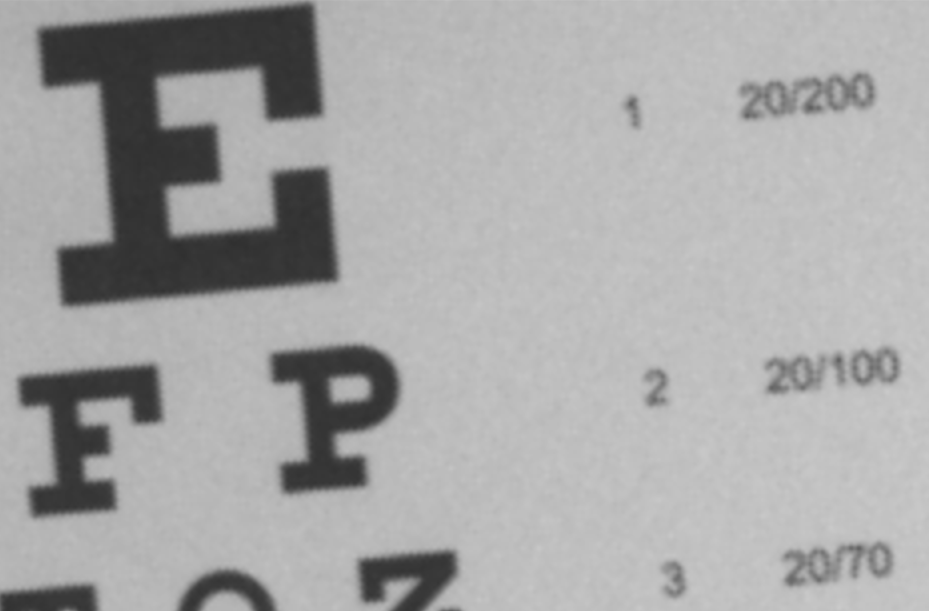}
    \caption{Sample from the first camera.}
  \end{subfigure}\hfill
  \begin{subfigure}[t]{0.48\linewidth}
    \includegraphics[width=\linewidth]{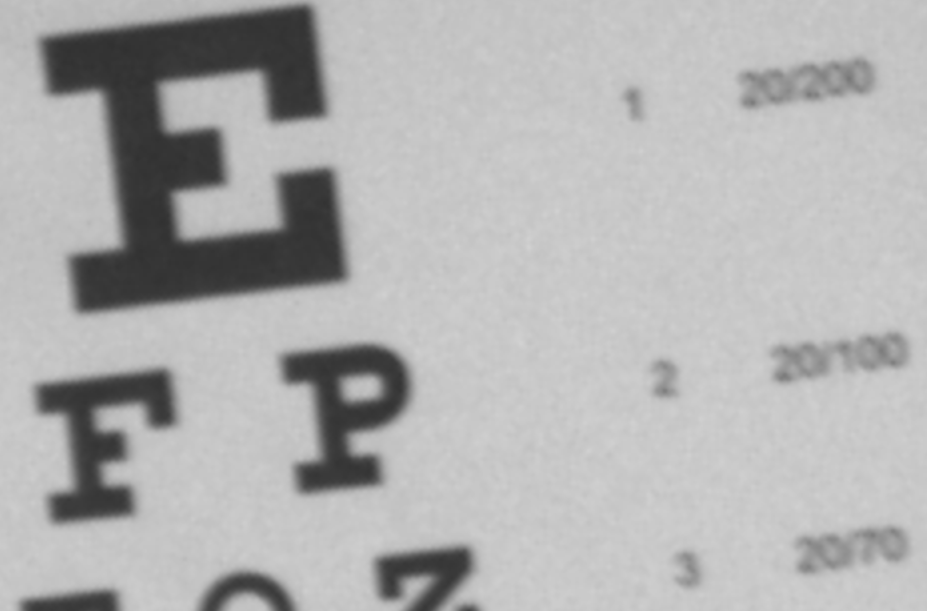}
    \caption{Sample from the second camera.}
  \end{subfigure}

  \begin{subfigure}[t]{0.60\linewidth}
    \includegraphics[width=\linewidth]{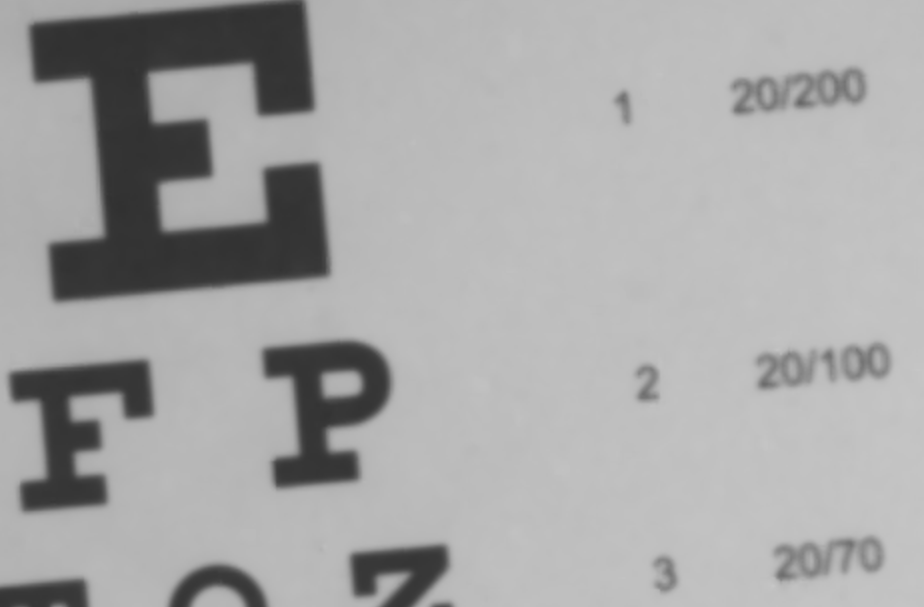}
    \caption{SR pipeline output with AVC.}
  \end{subfigure}
  \caption{Visual-test samples and SR output for the letter-pattern region. The output pixel count is four times the input pixel count.}
  \label{fig:visual-letters}
\end{figure}

\begin{figure}[htbp]
  \centering
  \begin{subfigure}[t]{0.48\linewidth}
    \includegraphics[width=\linewidth]{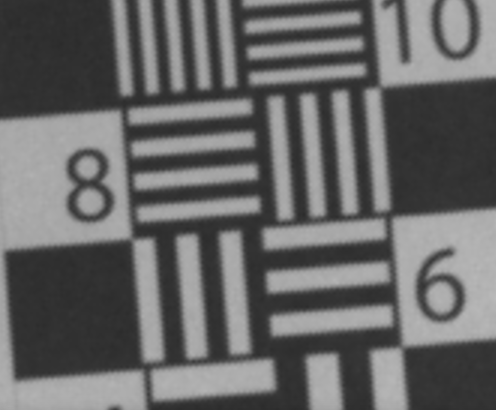}
    \caption{Sample from the first camera.}
  \end{subfigure}\hfill
  \begin{subfigure}[t]{0.48\linewidth}
    \includegraphics[width=\linewidth]{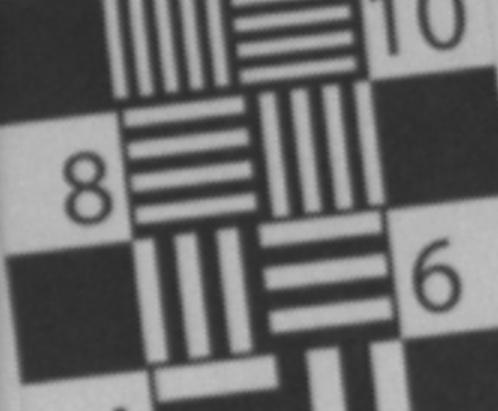}
    \caption{Sample from the second camera.}
  \end{subfigure}

  \begin{subfigure}[t]{0.60\linewidth}
    \includegraphics[width=\linewidth]{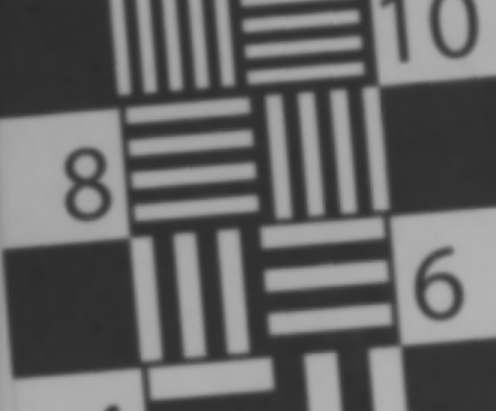}
    \caption{SR pipeline output with AVC.}
  \end{subfigure}
  \caption{Visual-test samples and SR output for the line-pattern region. The output pixel count is four times the input pixel count.}
  \label{fig:visual-lines}
\end{figure}

Visual inspection complements the quantitative evaluation. Figures~\ref{fig:visual-letters} and~\ref{fig:visual-lines} show geometric-pattern regions containing one sample from each camera, selected from the 16 images per camera in a burst, and the corresponding SR output obtained with AVC active. No metric-based comparison is performed for these examples, so the burst images are not resized before entering the pipeline. The input images have a resolution of $1024\times1024$ and the outputs have a resolution of $2048\times2048$, so each output contains four times as many pixels as an input. Across circular and linear features, the output examples exhibit sharper edges, finer structural details, and less noise than the raw inputs.

\section{Experimental Evaluation}
\label{sec:evaluation}

The evaluation connects the two experimental paths defined in
Sec.~\ref{sec:experimental-protocol}: inertial measurements establish the effect
of active vibration control (AVC) on the structure, and visual and metric-based
tests establish how that controlled condition affects the super-resolution (SR)
output.  The evidence is reported for the laboratory platform and the three
recorded scenarios defined in the protocol; the additional analyses below test
the robustness of that evidence without extending the claim beyond the
experiment.

\subsection{Inertial/control performance}

Using the weight functions in Sec.~\ref{ch:AVC}, the minimum value found by the
Riccati-based solver is $\gamma=23.12$.  The controller has a multiloop gain
margin of $3.34$~dB and a multiloop phase margin of $21.5\degree$ at the critical
frequency.  A Monte Carlo analysis of 500 perturbed plant variants evaluates
robustness: natural frequencies vary by $\pm10\%$, damping ratios by $\pm15\%$,
and the actuator-gain parameter by $\pm60\%$ to represent the mismatch observed
experimentally.

\begin{figure}[htbp]
  \centering
  \includegraphics[width=0.72\linewidth]{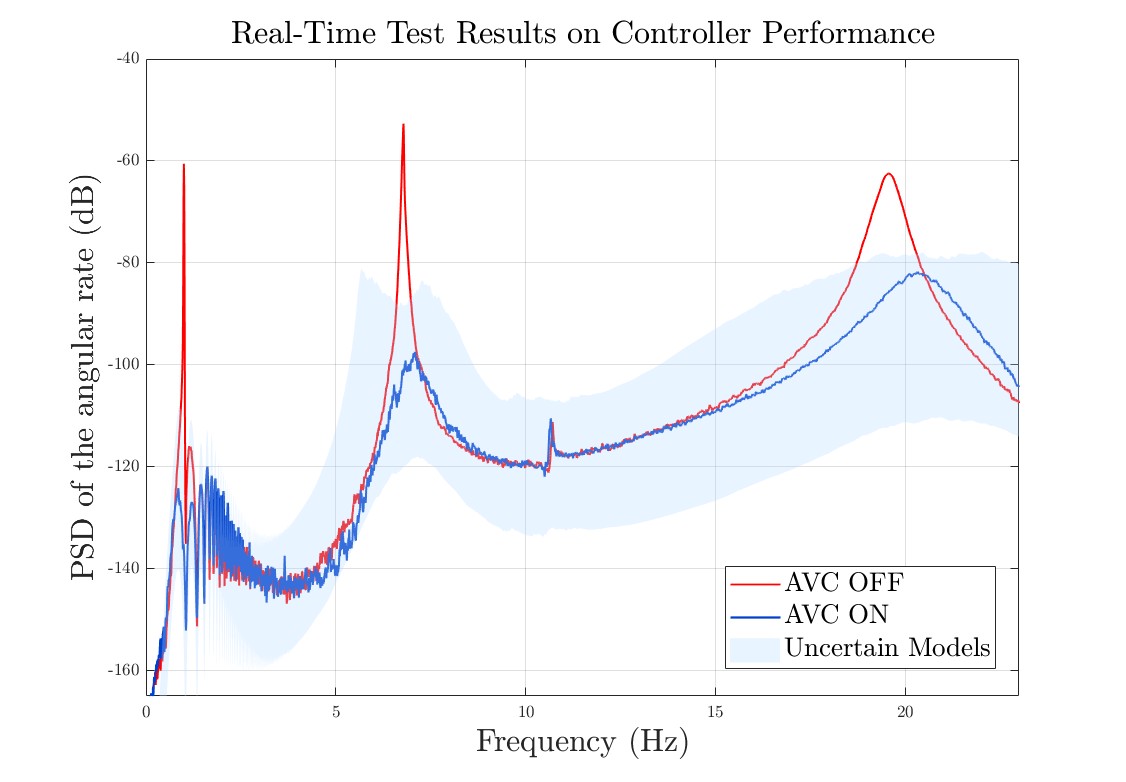}
  \caption{Control performance at the second camera location in the frequency
  domain. The open-loop case without vibration control is compared with the
  case in which AVC is enabled. The shaded envelope is the stable closed-loop
  response domain obtained from the Monte Carlo analysis.}
  \label{fig:control-performance}
\end{figure}

The controller was implemented on an embedded computer for real-time
closed-loop tests.  Figure~\ref{fig:control-performance} presents the
frequency-domain power spectral density (PSD) over the control bandwidth.  The
closed-loop responses remain stable throughout the considered uncertainty
envelope.  Low-frequency vibration associated with the first two vertical
bending modes is substantially attenuated, and the third vertical bending mode
is moderately suppressed.  Some amplification above 20~Hz is consistent with
the Bode sensitivity integral.  The 10.67~Hz torsional mode is unaffected
because the present actuators exert torque mainly about the $y$-axis; changing
their placement or alignment would change the controllable modal directions.
This inertial result provides the control-side evidence for the accepted claim,
within the modeled uncertainty and tested bandwidth.

\subsection{Visual SR output}

The visual-test protocol and its canonical examples appear in
Figs.~\ref{fig:visual-letters} and~\ref{fig:visual-lines}.  The letter-pattern
figure is the version retained from the accepted manuscript; the line-pattern
figure adds the distinct example supplied with the supplementary evaluation.
In both cases, the AVC-enabled SR output has four times the input pixel count
and exhibits sharper edges, finer structural detail, and less visible noise
than the two raw camera samples.  These examples support the image-quality
interpretation of the accepted result, but are qualitative and do not by
themselves establish performance outside the recorded laboratory scenes.

\subsection{Quantitative ground-truth construction}

\begin{figure}[htbp]
  \centering
  \begin{subfigure}[t]{0.32\linewidth}
    \includegraphics[width=\linewidth]{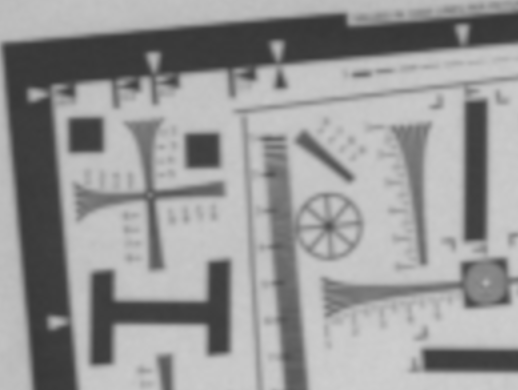}
    \caption{Burst sample.}
  \end{subfigure}\hfill
  \begin{subfigure}[t]{0.32\linewidth}
    \includegraphics[width=\linewidth]{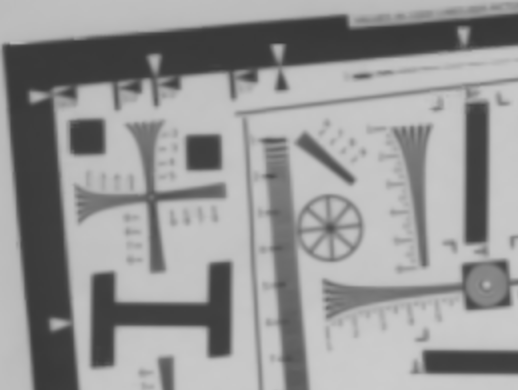}
    \caption{SR output.}
  \end{subfigure}\hfill
  \begin{subfigure}[t]{0.32\linewidth}
    \includegraphics[width=\linewidth]{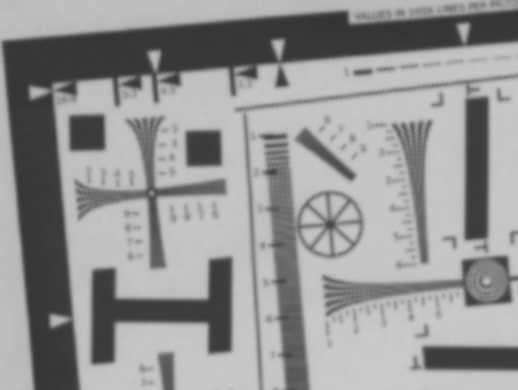}
    \caption{Ground truth.}
  \end{subfigure}
  \caption{Images used in the quantitative evaluation: a sample from a burst,
  the SR-pipeline output, and the ground-truth image used for the nine-metric
  comparison. The visible improvement from the burst sample to the output is
  supplemented by comparison with the ground truth.}
  \label{fig:quantitative-comparison}
\end{figure}

The quantitative procedure in Sec.~\ref{sec:experimental-protocol} compares a
$1024\times1024$ SR output with an aligned, brightness-adjusted
$1024\times1024$ ground-truth (GT) window after the burst inputs have been
resized to $512\times512$.  Figure~\ref{fig:quantitative-comparison} shows the
three images used in one such comparison.  The construction enables a
laboratory metric evaluation despite the absence of a separate high-resolution
reference sensor; the resizing and region-of-interest selection are evaluation
operations, not parts of the normal SR workflow.

\subsection{Nine-metric summary}

Table~\ref{tab:metric-summary} retains the accepted manuscript's complete
nine-metric summary.  MAE, RMSE, PSNR, and SSIM provide established pixel and
structural measures, while DISTS, ERQA, FSIM, MS-SSIM, and PieAPP add edge,
texture, and perceptual measures.  Across these complementary measures, the
with-AVC results improve over the without-AVC results and, in several cases,
approach the disturbance-free values.  This agreement with the inertial result
in Fig.~\ref{fig:control-performance} supports the accepted conclusion that AVC
improves SR image quality on the demonstrated platform.

\begin{table}[htbp]
  \centering
  \caption{Averaged image-quality results for the three test scenarios. Results
  are mean $\pm$ standard deviation. Arrows indicate whether lower
  ($\downarrow$) or higher ($\uparrow$) values are preferred.}
  \label{tab:metric-summary}
  \resizebox{\textwidth}{!}{%
  \begin{tabular}{@{}lccclccc@{}}
    \toprule
     & \textbf{Disturbance-free} & \textbf{Without AVC} & \textbf{With AVC} &
     & \textbf{Disturbance-free} & \textbf{Without AVC} & \textbf{With AVC} \\
    \midrule
    \textbf{MAE} [$\downarrow$] & $0.006 \pm 4.29e$-$4$ & $0.036 \pm 3.05e$-$3$ & $0.011 \pm 5.89e$-$4$
      & \textbf{MS-SSIM} [$\uparrow$] & $0.994 \pm 9.81e$-$4$ & $0.927 \pm 2.00e$-$2$ & $0.990 \pm 7.63e$-$4$ \\
    \textbf{RMSE} [$\downarrow$] & $0.014 \pm 1.88e$-$3$ & $0.091 \pm 6.79e$-$3$ & $0.028 \pm 7.69e$-$3$
      & \textbf{DISTS} [$\downarrow$] & $0.090 \pm 6.64e$-$3$ & $0.196 \pm 1.89e$-$2$ & $0.087 \pm 1.38e$-$2$ \\
    \textbf{PSNR} [$\uparrow$] & $36.8 \pm 8.62e$-$1$ & $20.8 \pm 6.50e$-$1$ & $31.5 \pm 2.60$
      & \textbf{FSIM} [$\uparrow$] & $0.996 \pm 1.73e$-$3$ & $0.894 \pm 2.21e$-$2$ & $0.990 \pm 1.15e$-$3$ \\
    \textbf{SSIM} [$\uparrow$] & $0.973 \pm 2.01e$-$3$ & $0.849 \pm 2.40e$-$2$ & $0.972 \pm 3.12e$-$3$
      & \textbf{PieAPP} [$\downarrow$] & $0.567 \pm 3.18e$-$2$ & $3.009 \pm 2.75e$-$1$ & $0.612 \pm 5.88e$-$2$ \\
    \textbf{ERQA} [$\uparrow$] & $0.861 \pm 6.74e$-$3$ & $0.434 \pm 7.43e$-$2$ & $0.842 \pm 1.23e$-$2$ & & & & \\
    \bottomrule
  \end{tabular}}
\end{table}

\subsection{Per-burst metric distributions}

\begin{figure}[htbp]
  \centering
  \begin{subfigure}[t]{0.32\linewidth}
    \includegraphics[width=\linewidth]{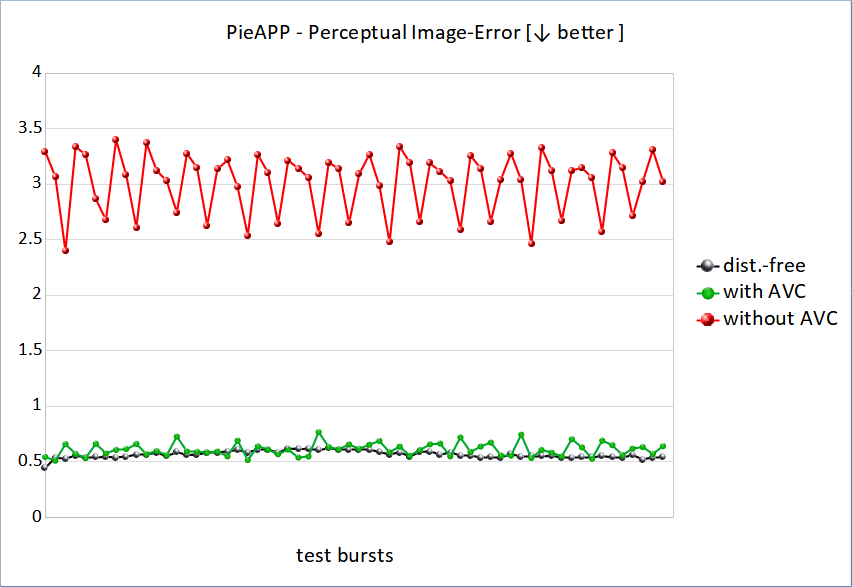}
    \caption{PieAPP}
  \end{subfigure}\hfill
  \begin{subfigure}[t]{0.32\linewidth}
    \includegraphics[width=\linewidth]{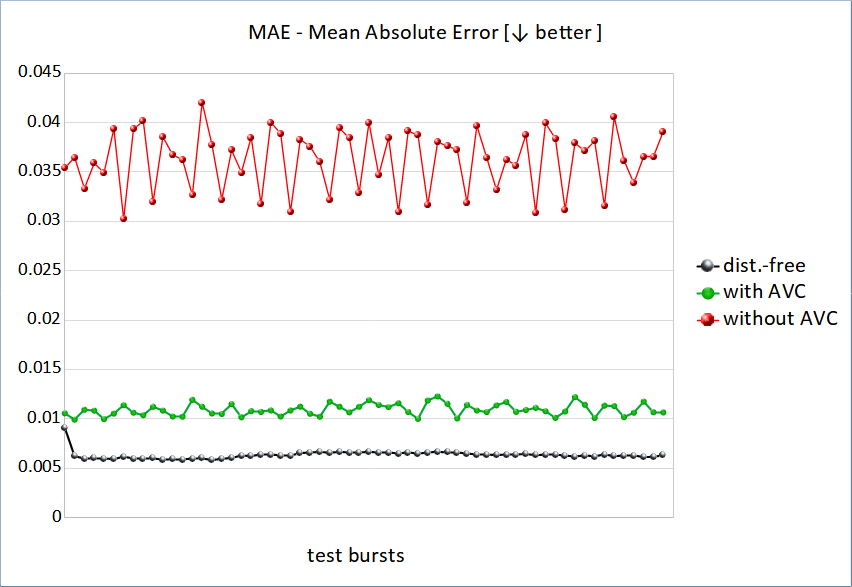}
    \caption{MAE}
  \end{subfigure}\hfill
  \begin{subfigure}[t]{0.32\linewidth}
    \includegraphics[width=\linewidth]{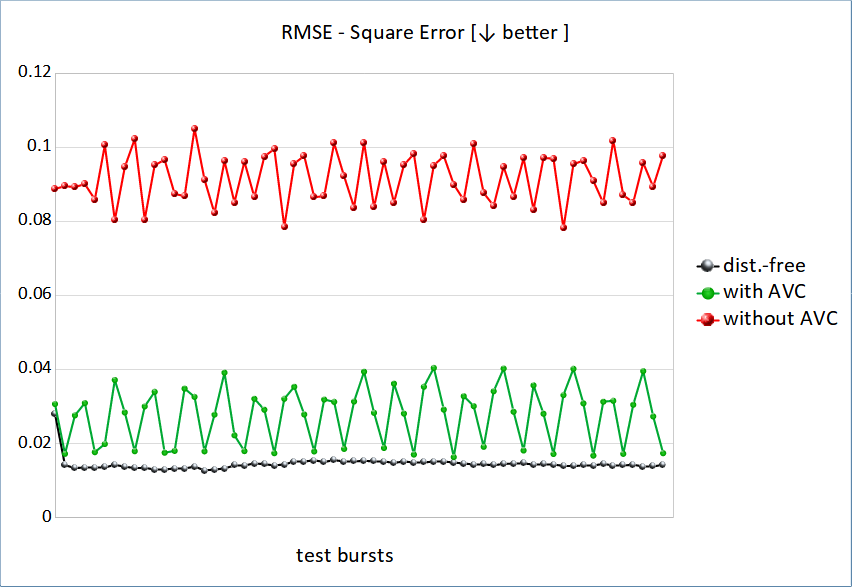}
    \caption{RMSE}
  \end{subfigure}

  \begin{subfigure}[t]{0.32\linewidth}
    \includegraphics[width=\linewidth]{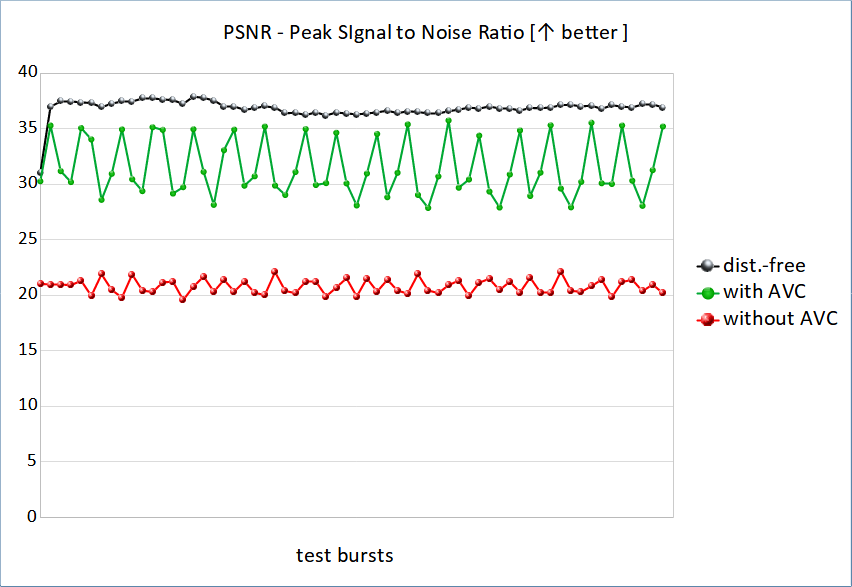}
    \caption{PSNR}
  \end{subfigure}\hfill
  \begin{subfigure}[t]{0.32\linewidth}
    \includegraphics[width=\linewidth]{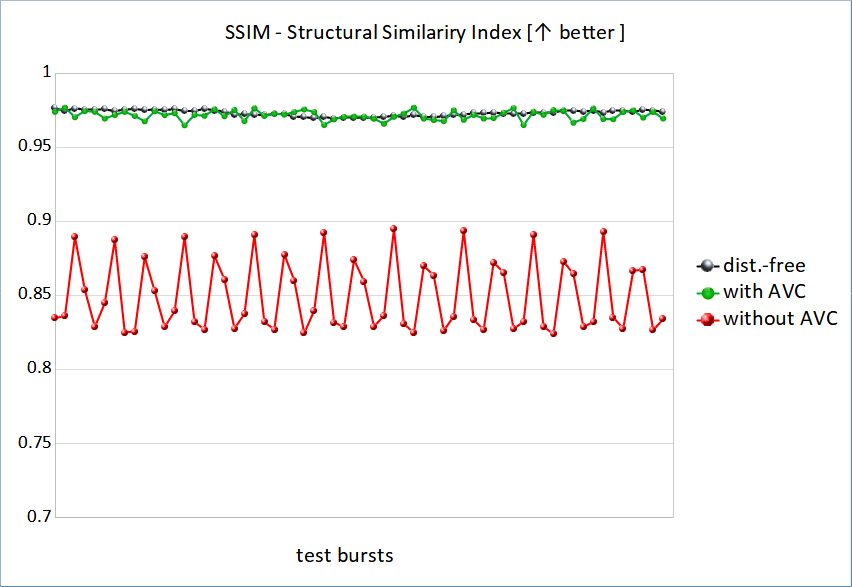}
    \caption{SSIM}
  \end{subfigure}\hfill
  \begin{subfigure}[t]{0.32\linewidth}
    \includegraphics[width=\linewidth]{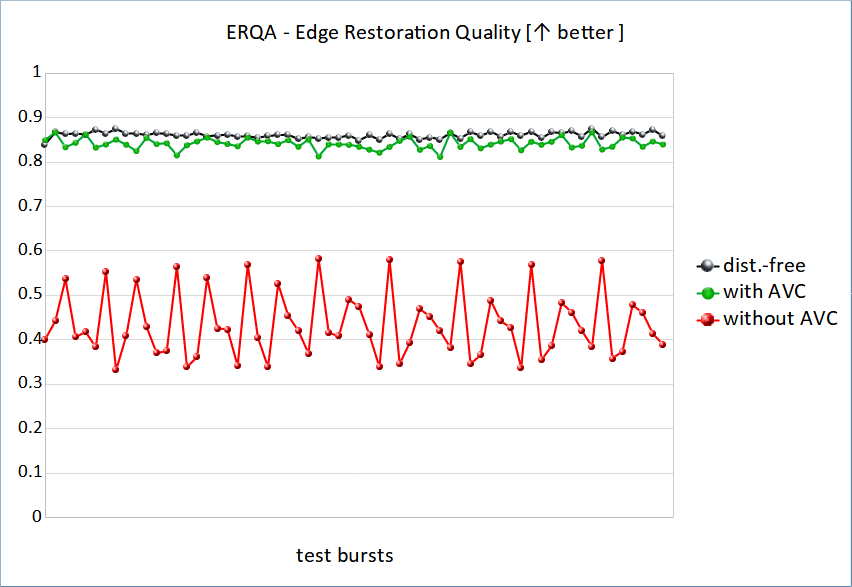}
    \caption{ERQA}
  \end{subfigure}

  \begin{subfigure}[t]{0.32\linewidth}
    \includegraphics[width=\linewidth]{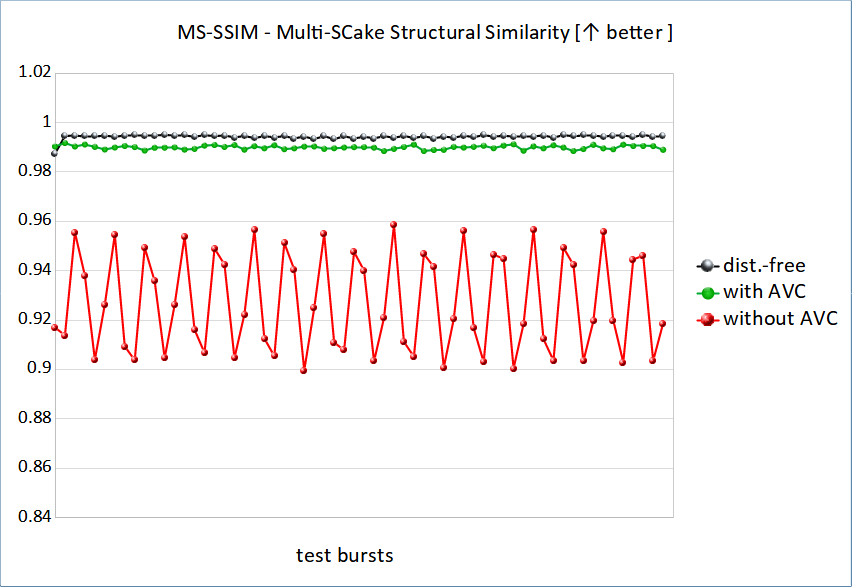}
    \caption{MS-SSIM}
  \end{subfigure}\hfill
  \begin{subfigure}[t]{0.32\linewidth}
    \includegraphics[width=\linewidth]{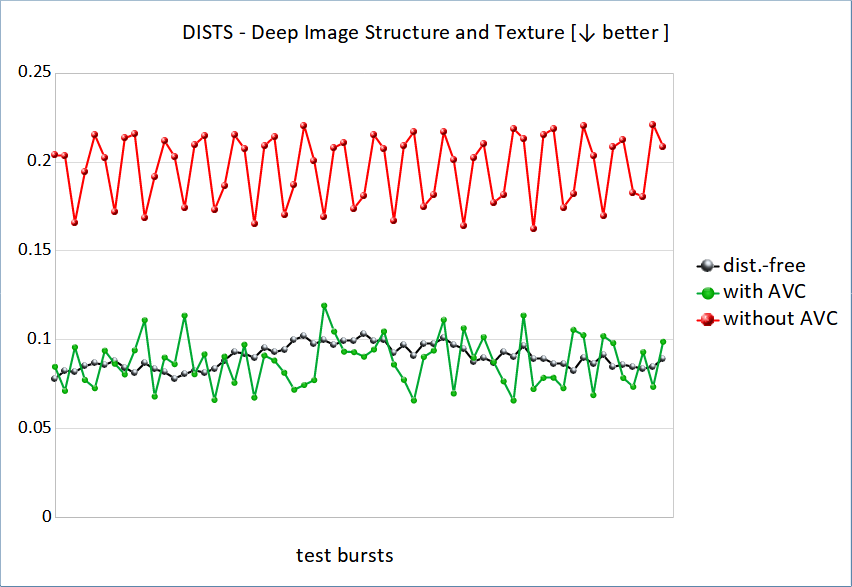}
    \caption{DISTS}
  \end{subfigure}\hfill
  \begin{subfigure}[t]{0.32\linewidth}
    \includegraphics[width=\linewidth]{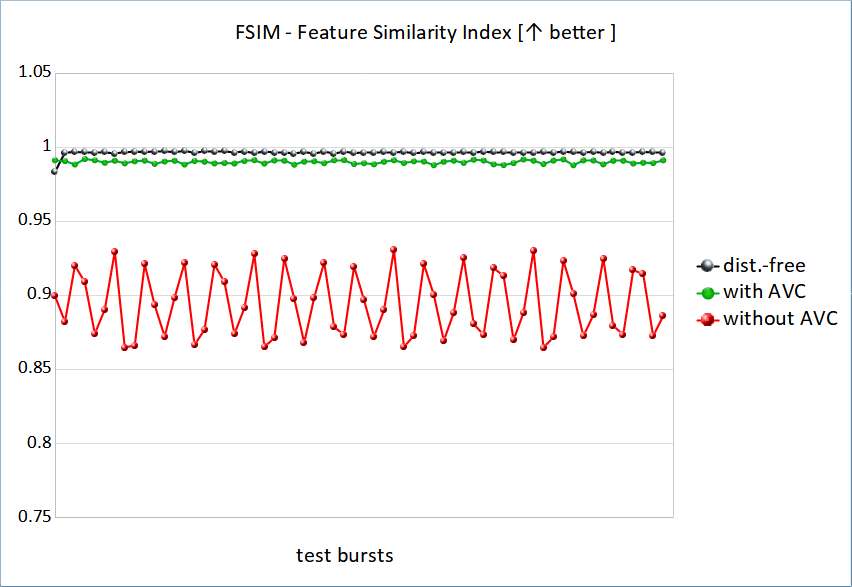}
    \caption{FSIM}
  \end{subfigure}
  \caption{Per-burst SR quality for all nine metrics in the disturbance-free,
  with-AVC, and without-AVC scenarios. Arrows in the plots indicate whether
  lower or higher values correspond to better quality.}
  \label{fig:metric-distributions}
\end{figure}

Figure~\ref{fig:metric-distributions} retains the accepted PieAPP plot and adds
the eight supplement-only plots, so every metric in
Table~\ref{tab:metric-summary} can be inspected burst by burst.  Each point is
the result for one burst.  The disturbance-free results generally have the
highest quality, the with-AVC results are close to them, and the without-AVC
results have lower average quality.  Their spread reflects, among other test
conditions, the phase of the $6.78$~Hz vibration at which an unsynchronized
burst begins.  Showing the individual bursts verifies that the accepted mean
result is not an artifact of reporting only an aggregate.

\subsection{Statistical significance analysis}

The with-AVC and without-AVC samples each contain 62 independent bursts.  The
bursts are not necessarily paired, and equal variances are not assumed, so a
Welch t-test was applied separately to each metric.  For sample means
$\bar{X}_1$ and $\bar{X}_2$, standard deviations $\sigma_1$ and $\sigma_2$,
and sample sizes $N_1$ and $N_2$, the reported statistics use
\begin{equation}
t = \frac{\bar{X}_1 - \bar{X}_2}
         {\operatorname{SE}\!\left(\bar{X}_1 - \bar{X}_2\right)},
\end{equation}
where
\begin{equation}
\operatorname{SE}\!\left(\bar{X}_1 - \bar{X}_2\right)
=\sqrt{\frac{\sigma_1^{2}}{N_1}+\frac{\sigma_2^{2}}{N_2}},
\end{equation}
and
\begin{equation}
\mathrm{df}=\frac{\left(\sigma_1^2/N_1 + \sigma_2^2/N_2\right)^2}{
\left(\sigma_1^2/N_1\right)^2/(N_1-1)
+\left(\sigma_2^2/N_2\right)^2/(N_2-1)}.
\end{equation}

Table~\ref{tab:statistical-evaluation} reports the resulting t-statistics and
degrees of freedom.  The source analysis concludes with high confidence that
the with-AVC improvement is statistically significant for all nine metrics.
This test strengthens the accepted laboratory comparison by accounting for
sample variability; it does not establish performance for untested platforms
or environments.

For each of the three scenarios, 1000 images were recorded by each camera,
giving 6000 images in total.  With 16 images per camera in each burst, 62
independent bursts were formed per scenario from 992 images per camera, and no
frame was reused across bursts.  Lighting varied during acquisition.  The
cameras operated at 30~fps without synchronization to the $6.78$~Hz disturbance,
so bursts began and ended at different vibration phases while retaining the
consecutive-frame changes used by the SR algorithm.

\begin{table}[htbp]
  \centering
  \caption{Summary of t-statistics and degrees of freedom (df) for the nine
  evaluation metrics.}
  \label{tab:statistical-evaluation}
  \resizebox{\textwidth}{!}{%
  \begin{tabular}{@{}lccccccccc@{}}
    \toprule
     & MAE & RMSE & PSNR & SSIM & ERQA & MS-SSIM & DISTS & FSIM & PieAPP \\
    \midrule
    $t$  & -64.131 & -48.997 & 31.574 & 39.745 & 42.616 & 24.862 & -36.617 & 34.163 & -67.163 \\
    $df$ & 65.526 & 120.151 & 68.593 & 63.061 & 64.324 & 61.177 & 111.713 & 61.330 & 66.572 \\
    \bottomrule
  \end{tabular}}
\end{table}

\subsection{Ground-truth sensitivity analysis}

\begin{figure}[htbp]
  \centering
  \begin{subfigure}[t]{0.32\linewidth}
    \includegraphics[height=0.20\textheight]{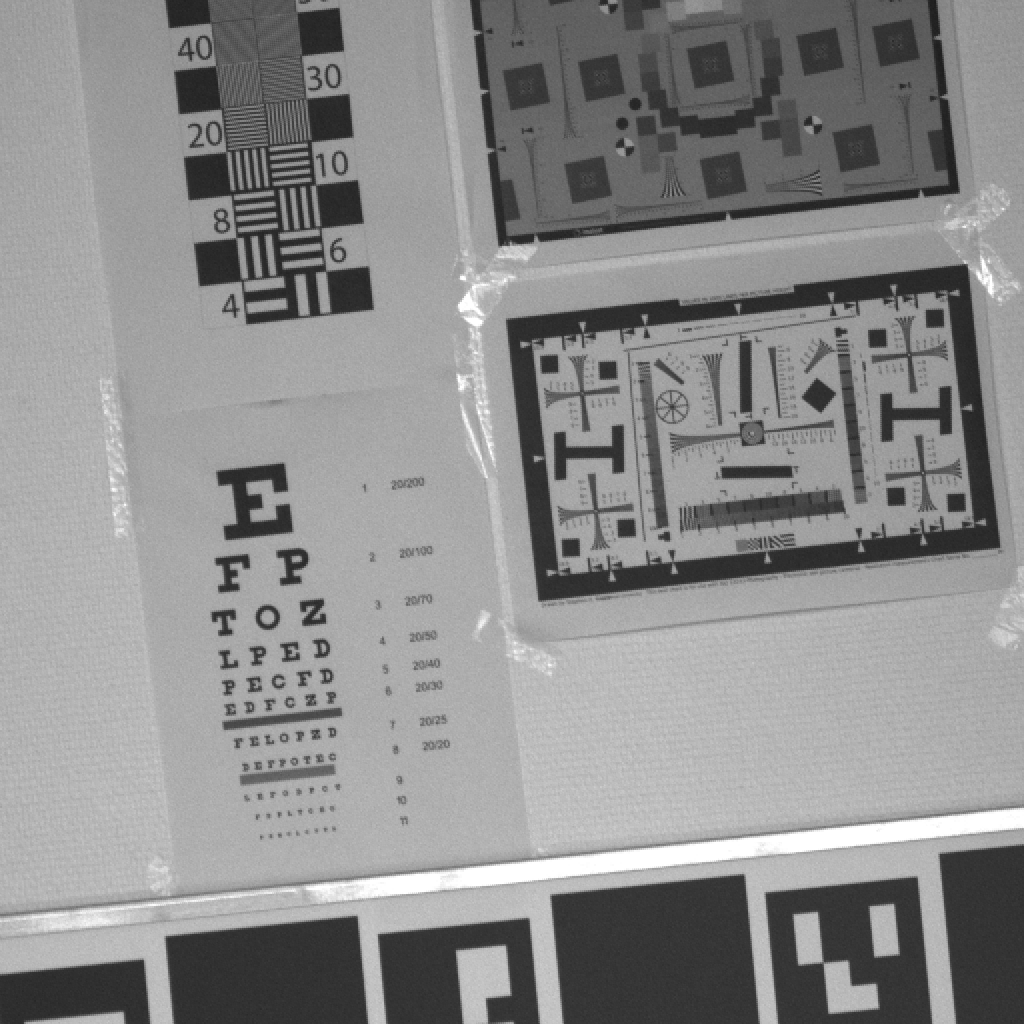}
    \caption{Sample 1}
  \end{subfigure}\hfill
  \begin{subfigure}[t]{0.32\linewidth}
    \includegraphics[height=0.20\textheight]{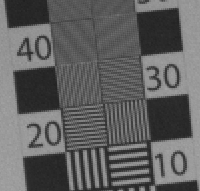}
    \caption{Sample 1, detail 1}
  \end{subfigure}\hfill
  \begin{subfigure}[t]{0.32\linewidth}
    \includegraphics[height=0.20\textheight]{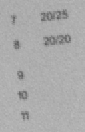}
    \caption{Sample 1, detail 2}
  \end{subfigure}

  \begin{subfigure}[t]{0.32\linewidth}
    \includegraphics[height=0.20\textheight]{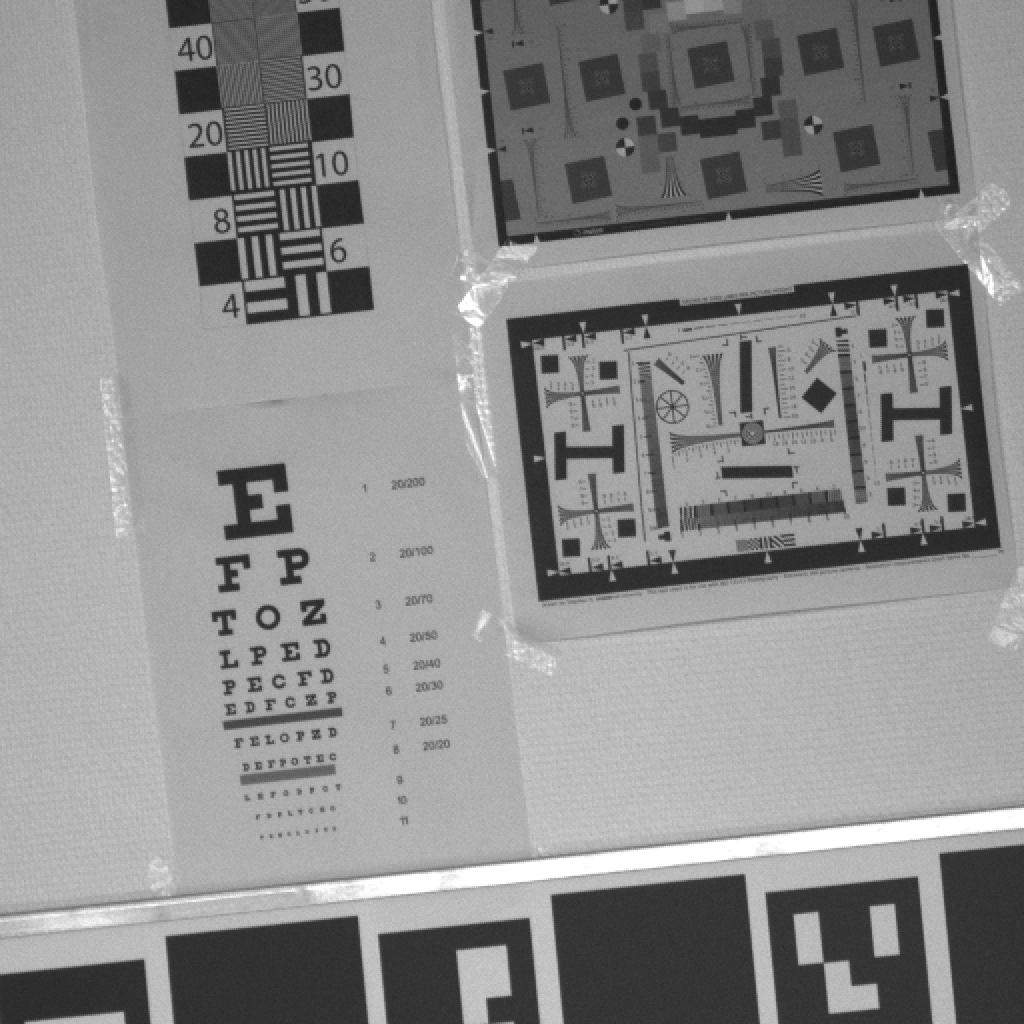}
    \caption{Sample 2}
  \end{subfigure}\hfill
  \begin{subfigure}[t]{0.32\linewidth}
    \includegraphics[height=0.20\textheight]{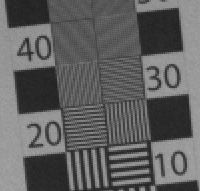}
    \caption{Sample 2, detail 1}
  \end{subfigure}\hfill
  \begin{subfigure}[t]{0.32\linewidth}
    \includegraphics[height=0.20\textheight]{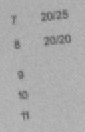}
    \caption{Sample 2, detail 2}
  \end{subfigure}

  \begin{subfigure}[t]{0.32\linewidth}
    \includegraphics[height=0.20\textheight]{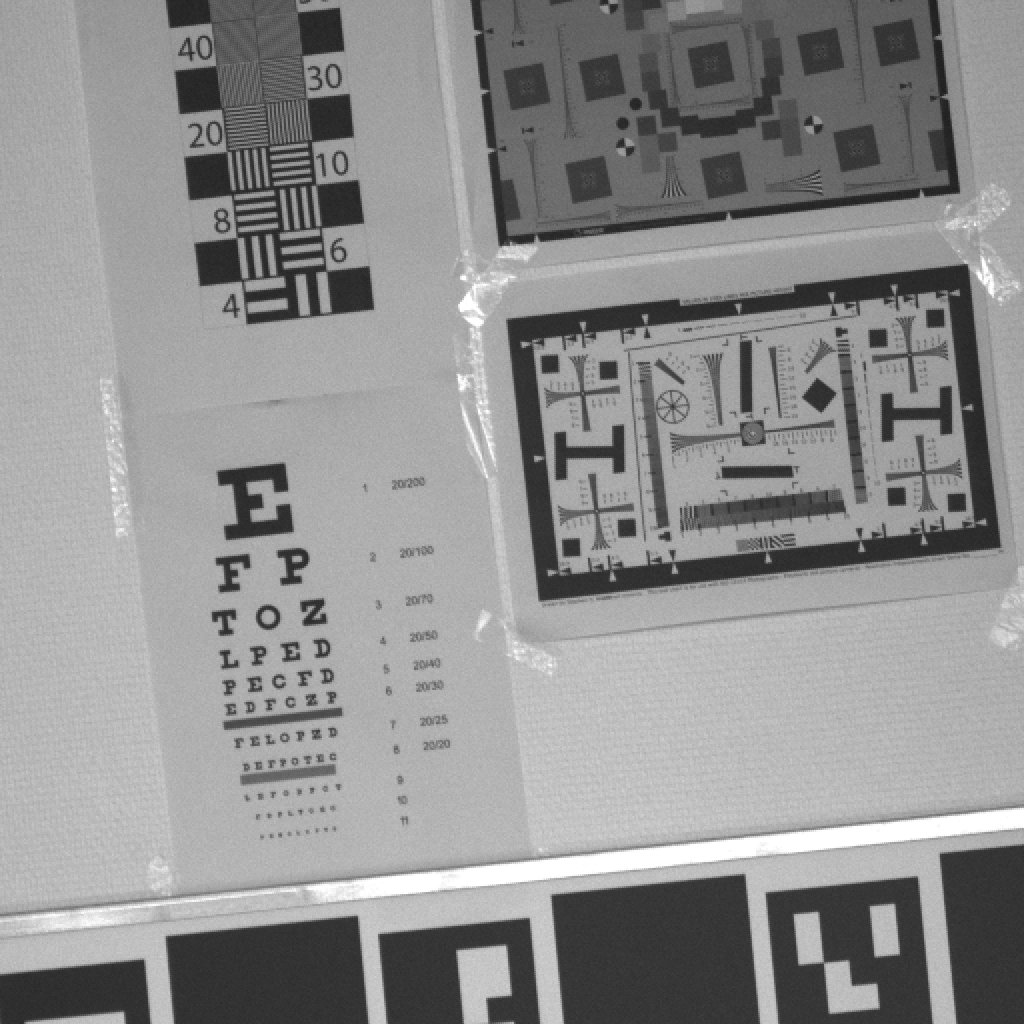}
    \caption{Sample 3}
  \end{subfigure}\hfill
  \begin{subfigure}[t]{0.32\linewidth}
    \includegraphics[height=0.20\textheight]{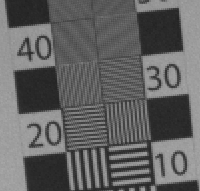}
    \caption{Sample 3, detail 1}
  \end{subfigure}\hfill
  \begin{subfigure}[t]{0.32\linewidth}
    \includegraphics[height=0.20\textheight]{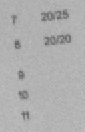}
    \caption{Sample 3, detail 2}
  \end{subfigure}
  \caption{Three input frames from the disturbance-free scenario and matched
  detail regions. Fine detail is visible, together with random noise in each
  frame.}
  \label{fig:ground-truth-input-frames}
\end{figure}

\begin{figure}[htbp]
  \centering
  \begin{subfigure}[t]{0.32\linewidth}
    \includegraphics[height=0.20\textheight]{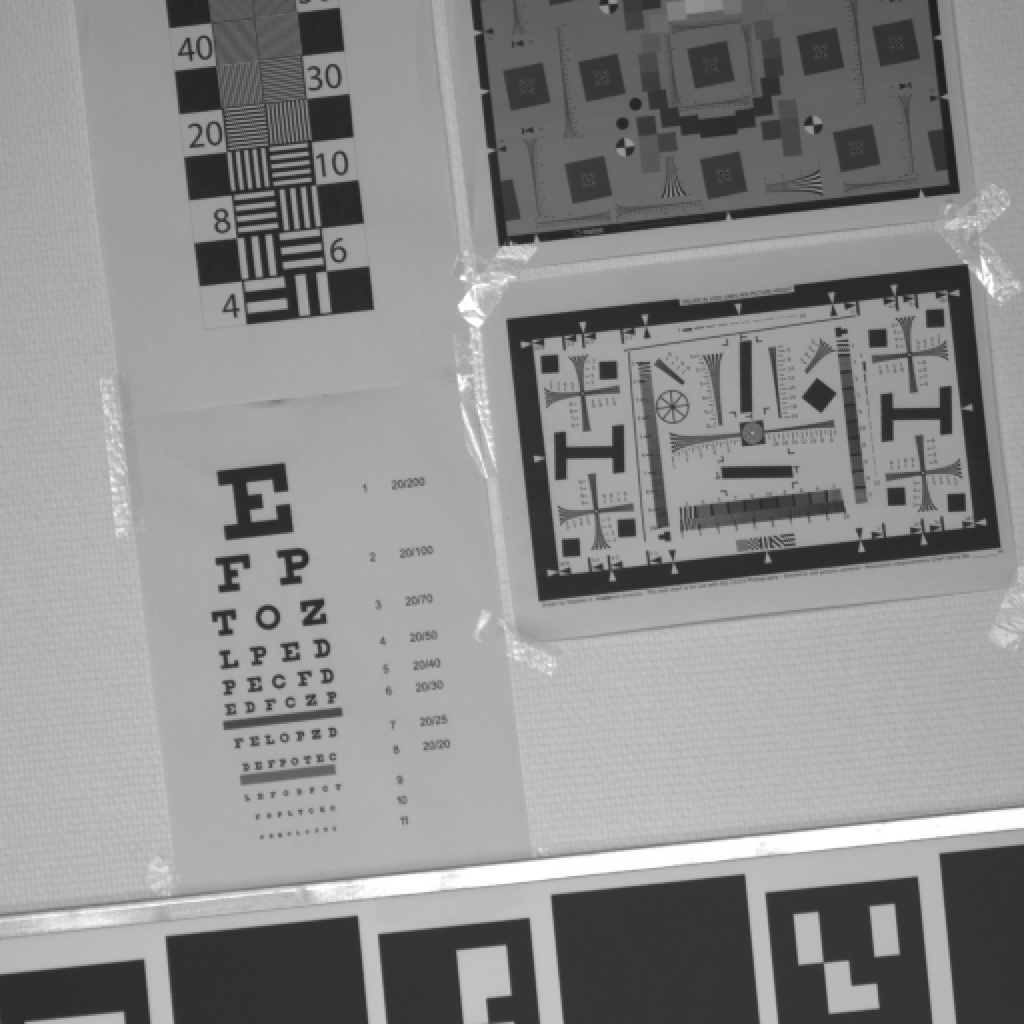}
    \caption{Ground truth}
  \end{subfigure}\hfill
  \begin{subfigure}[t]{0.32\linewidth}
    \includegraphics[height=0.20\textheight]{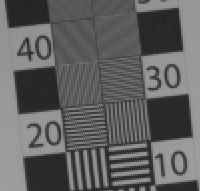}
    \caption{GT, detail 1}
  \end{subfigure}\hfill
  \begin{subfigure}[t]{0.32\linewidth}
    \includegraphics[height=0.20\textheight]{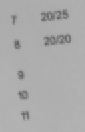}
    \caption{GT, detail 2}
  \end{subfigure}

  \begin{subfigure}[t]{0.32\linewidth}
    \includegraphics[height=0.20\textheight]{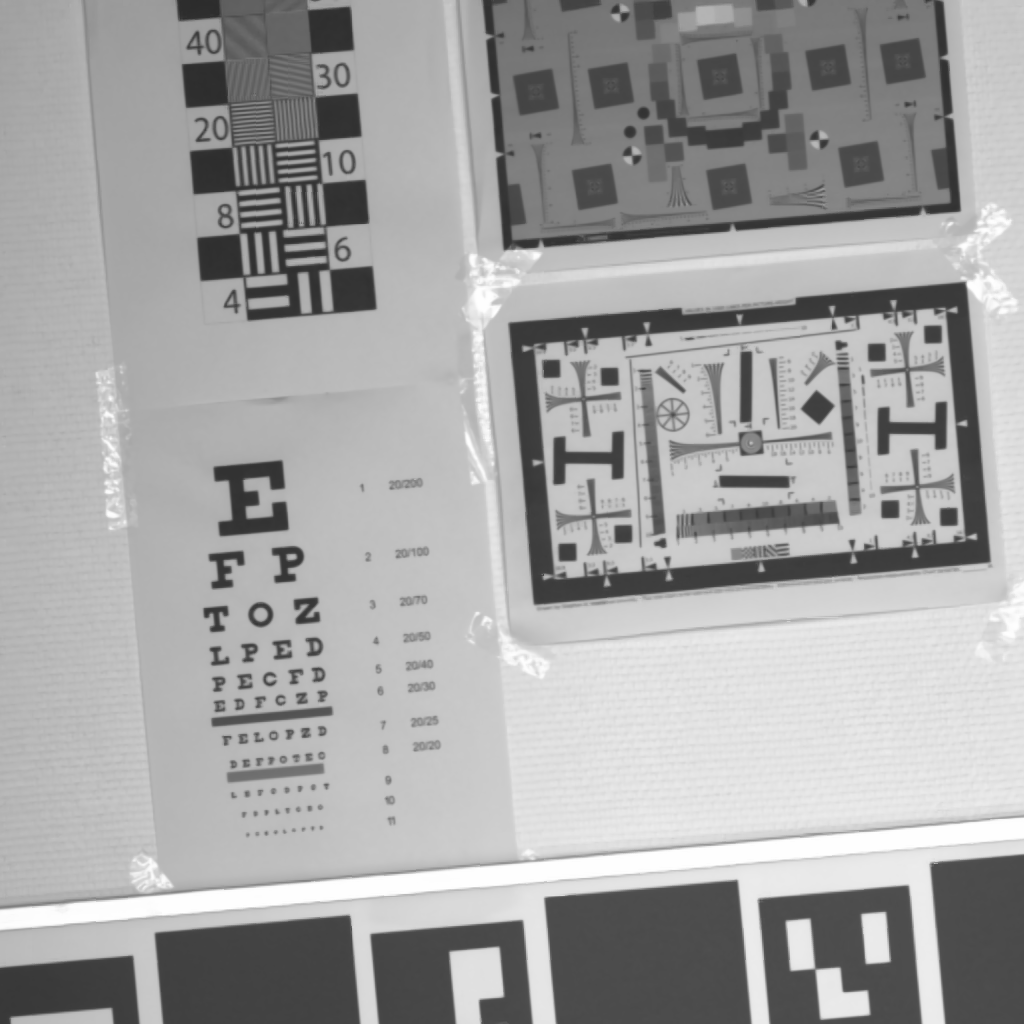}
    \caption{Output 1}
  \end{subfigure}\hfill
  \begin{subfigure}[t]{0.32\linewidth}
    \includegraphics[height=0.20\textheight]{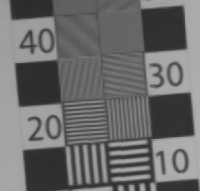}
    \caption{Output 1, detail 1}
  \end{subfigure}\hfill
  \begin{subfigure}[t]{0.32\linewidth}
    \includegraphics[height=0.20\textheight]{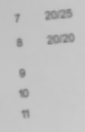}
    \caption{Output 1, detail 2}
  \end{subfigure}

  \begin{subfigure}[t]{0.32\linewidth}
    \includegraphics[height=0.20\textheight]{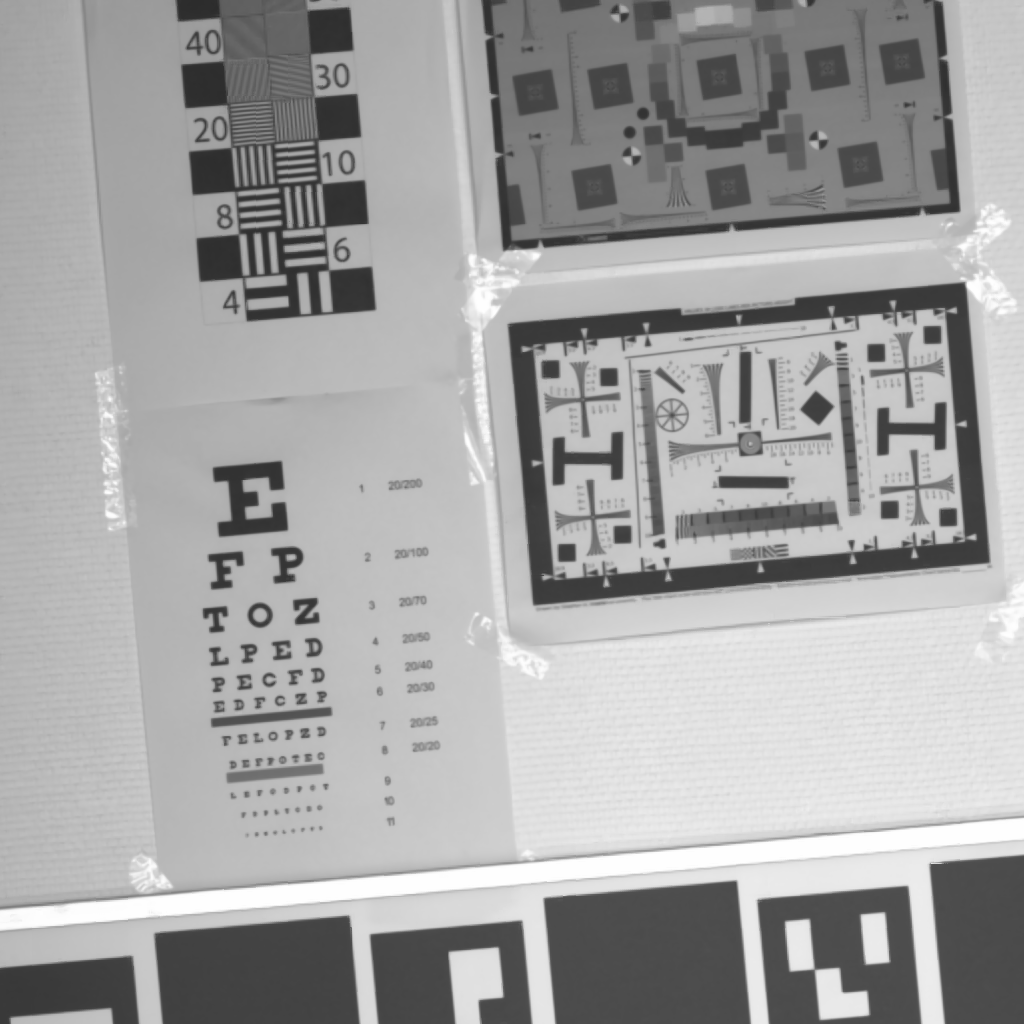}
    \caption{Output 2}
  \end{subfigure}\hfill
  \begin{subfigure}[t]{0.32\linewidth}
    \includegraphics[height=0.20\textheight]{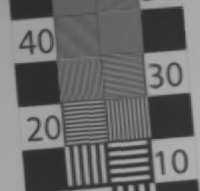}
    \caption{Output 2, detail 1}
  \end{subfigure}\hfill
  \begin{subfigure}[t]{0.32\linewidth}
    \includegraphics[height=0.20\textheight]{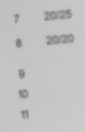}
    \caption{Output 2, detail 2}
  \end{subfigure}
  \caption{The averaged ground-truth image and two disturbance-free SR outputs,
  with matched detail regions. Comparison with
  Fig.~\ref{fig:ground-truth-input-frames} shows that averaging suppresses
  random noise while retaining high-frequency detail.}
  \label{fig:ground-truth-output-details}
\end{figure}

The fixed-camera frames used to construct the GT preserve high-frequency
detail while containing random noise.  Figure~\ref{fig:ground-truth-input-frames}
shows three such disturbance-free frames and matched details.  Their average
suppresses random noise without removing the visible fine structure, as shown
alongside two disturbance-free SR outputs in
Fig.~\ref{fig:ground-truth-output-details}.

\begin{figure}[htbp]
  \centering
  \begin{subfigure}[t]{0.32\linewidth}
    \includegraphics[width=\linewidth]{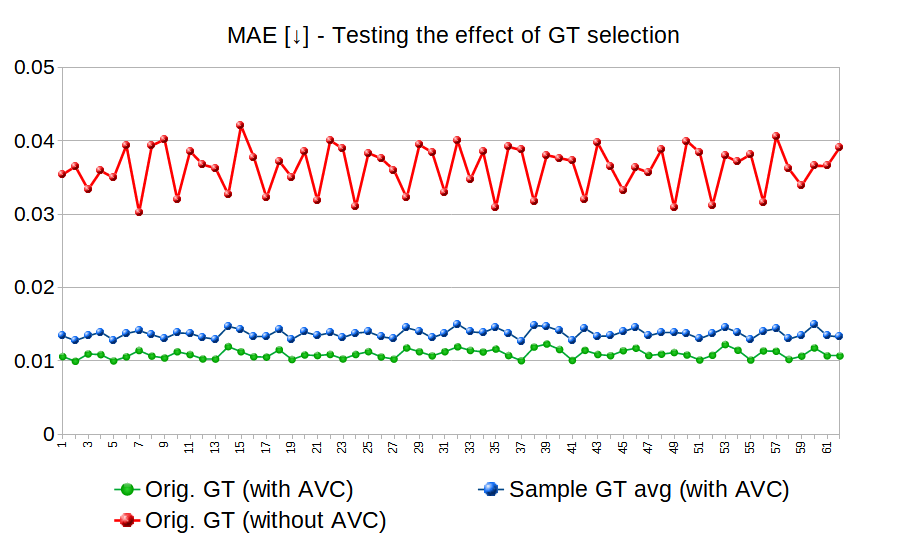}
    \caption{MAE}
  \end{subfigure}\hfill
  \begin{subfigure}[t]{0.32\linewidth}
    \includegraphics[width=\linewidth]{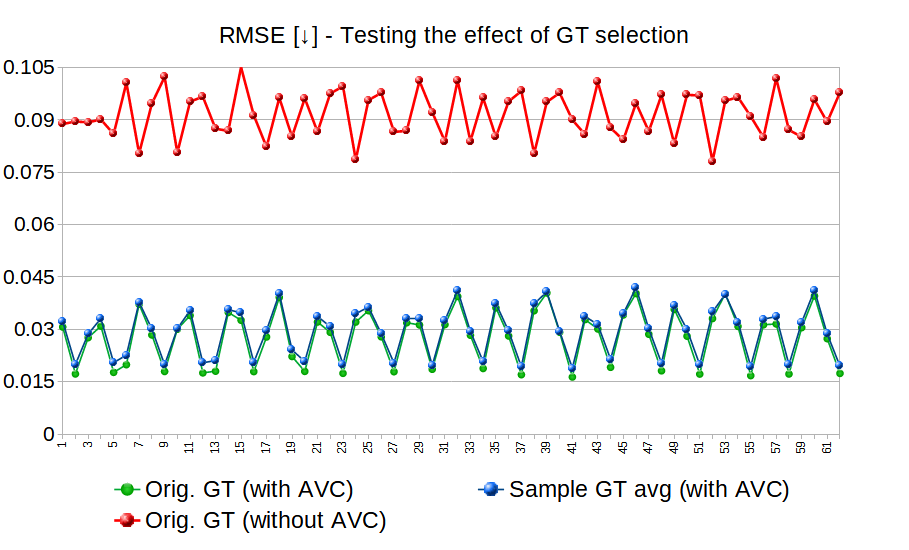}
    \caption{RMSE}
  \end{subfigure}\hfill
  \begin{subfigure}[t]{0.32\linewidth}
    \includegraphics[width=\linewidth]{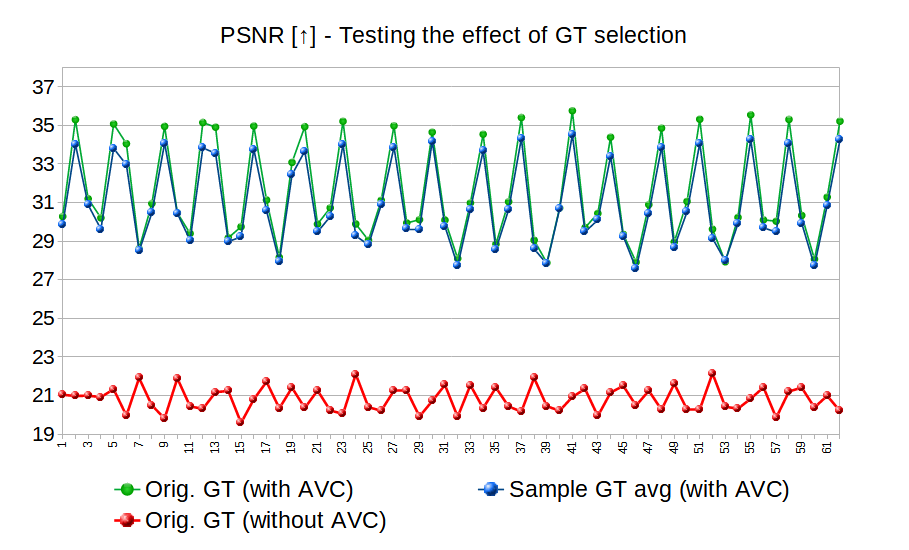}
    \caption{PSNR}
  \end{subfigure}

  \begin{subfigure}[t]{0.32\linewidth}
    \includegraphics[width=\linewidth]{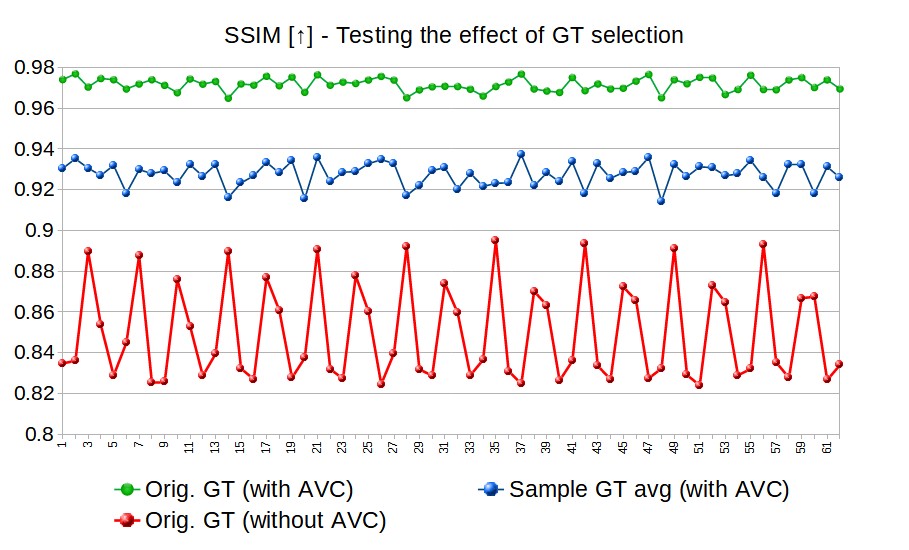}
    \caption{SSIM}
  \end{subfigure}\hfill
  \begin{subfigure}[t]{0.32\linewidth}
    \includegraphics[width=\linewidth]{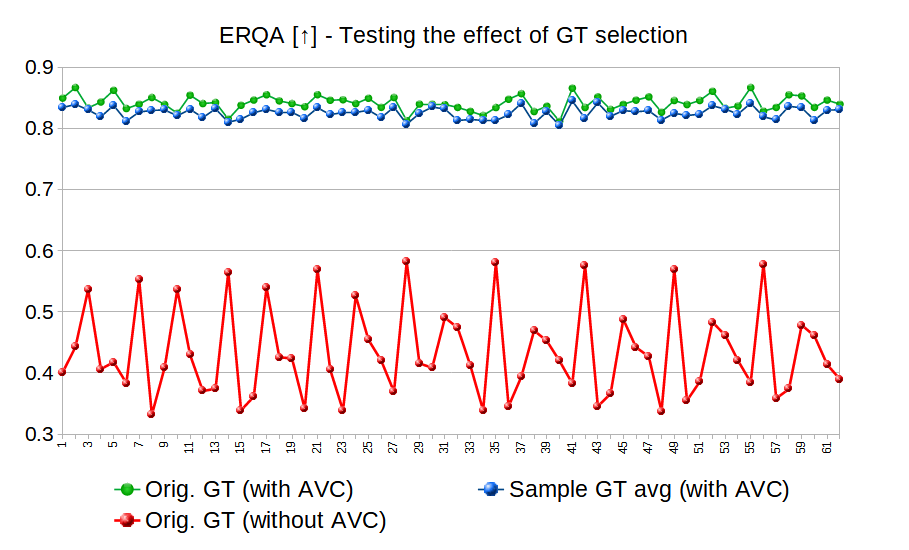}
    \caption{ERQA}
  \end{subfigure}\hfill
  \begin{subfigure}[t]{0.32\linewidth}
    \includegraphics[width=\linewidth]{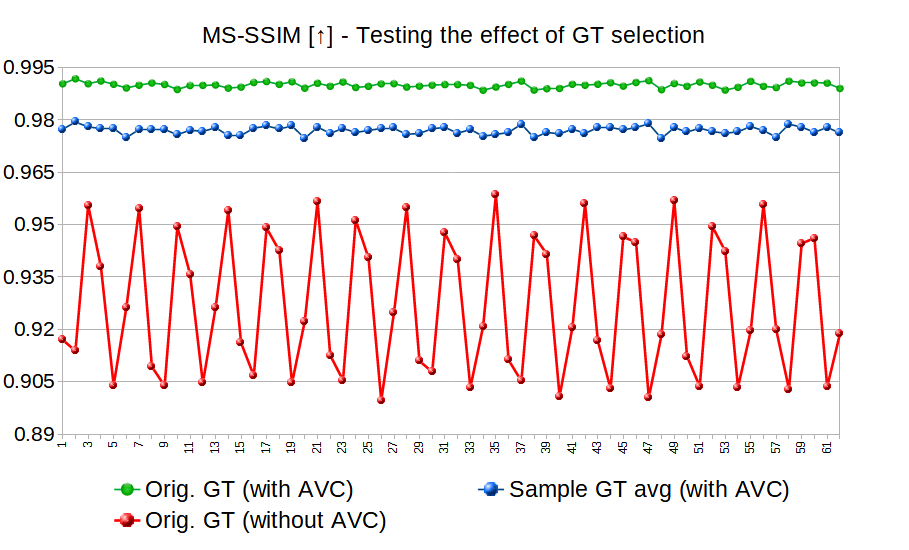}
    \caption{MS-SSIM}
  \end{subfigure}

  \begin{subfigure}[t]{0.32\linewidth}
    \includegraphics[width=\linewidth]{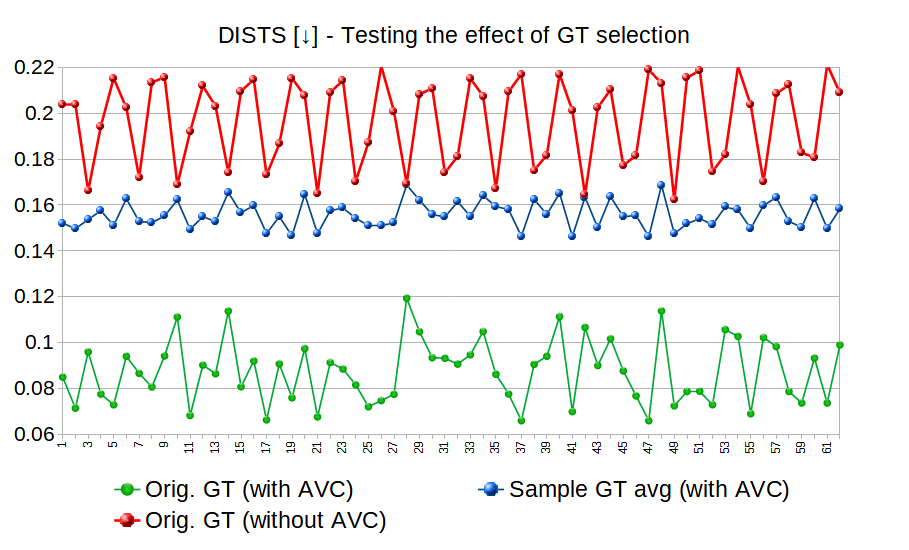}
    \caption{DISTS}
  \end{subfigure}\hfill
  \begin{subfigure}[t]{0.32\linewidth}
    \includegraphics[width=\linewidth]{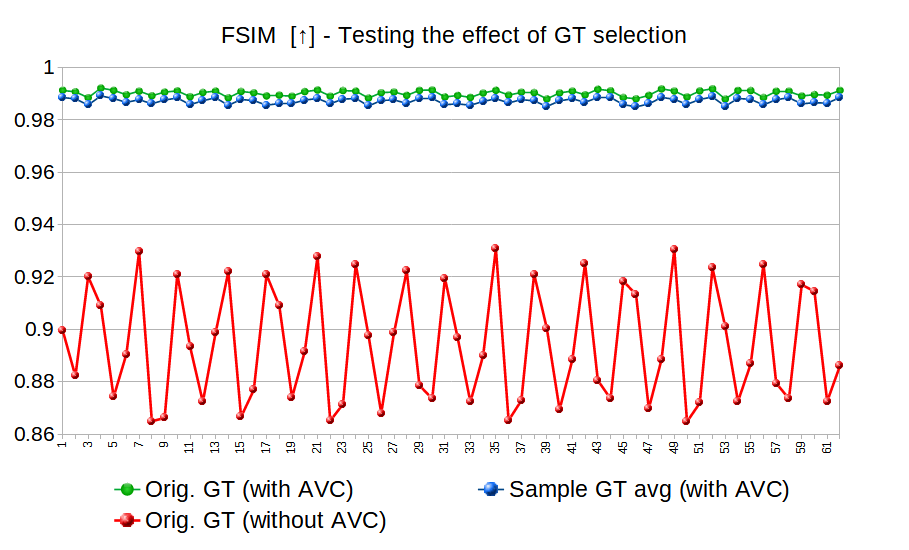}
    \caption{FSIM}
  \end{subfigure}\hfill
  \begin{subfigure}[t]{0.32\linewidth}
    \includegraphics[width=\linewidth]{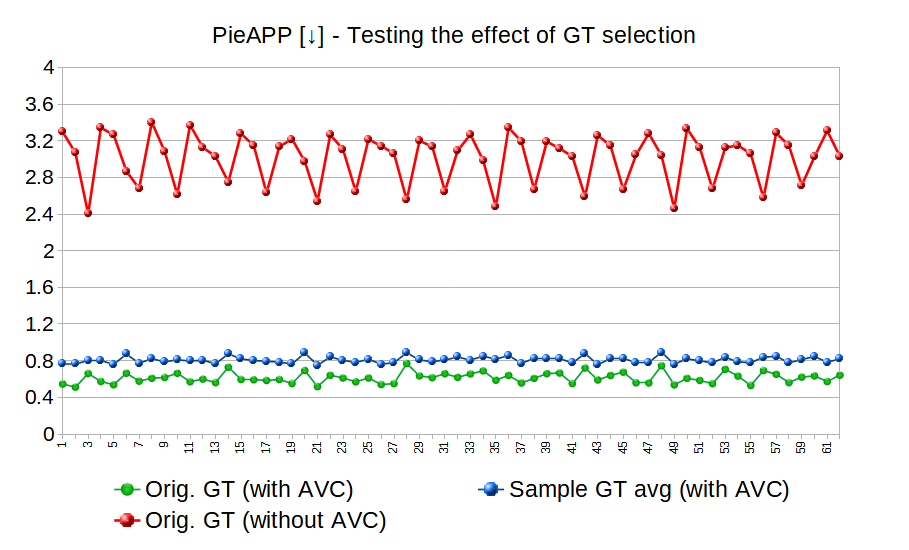}
    \caption{PieAPP}
  \end{subfigure}
  \caption{Sensitivity of the nine evaluation metrics to ground-truth
  construction. ``Sample GT avg.'' averages results from eleven individual
  candidate frames; ``Orig. GT'' uses the averaged ground truth. Original-GT
  without-AVC results are shown only to contextualize the differences. Arrows
  indicate whether lower or higher values correspond to better quality.}
  \label{fig:ground-truth-study}
\end{figure}

Metric sensitivity to GT selection was tested with eleven individual
disturbance-free frames selected at regular intervals.  Each frame was used as
GT for the same with-AVC SR outputs, and the eleven results were averaged per
burst.  Figure~\ref{fig:ground-truth-study} compares that average with the
original averaged GT and includes the original-GT without-AVC curves as a
reference for practical scale.  RMSE and PSNR are nearly insensitive to the GT
choice; MAE, ERQA, MS-SSIM, and FSIM change slightly; and SSIM, DISTS, and
PieAPP show more noticeable differences.  For MAE, for example, the alternative
and original GT curves differ but retain the same trend, and the difference is
small compared with the without-AVC reference.  Thus the sensitivity study
qualifies individual metric values while preserving the comparative AVC result
within the recorded data.

\clearpage
\subsection{Bicubic baseline comparison}

To check whether the observed improvement is specific to the selected burst SR
algorithm, a single-frame bicubic interpolation baseline was evaluated on the
same with-AVC and without-AVC data series.  One frame from each of the 62 bursts
was used as the bicubic input. The mean and standard deviation were then
calculated for the 62 output images using all nine metrics.

Bicubic interpolation does not fuse a burst, so this comparison removes burst
size and other multi-frame fusion choices while retaining the change in AVC
condition.  Table~\ref{tab:bicubic-baseline} shows improvement with AVC for all
nine metrics under both SR approaches.  The baseline therefore provides an
algorithm-independent check on the direction of the accepted AVC effect; it
does not claim that bicubic interpolation is equivalent to the proposed SR
pipeline or that either method has been validated outside this experiment.

\begin{table}[htbp]
  \centering
  \caption{Comparison of the without-AVC and with-AVC scenarios using the
  proposed SR pipeline and bicubic interpolation. Arrows indicate whether
  higher ($\uparrow$) or lower ($\downarrow$) values are preferred; results are
  mean $\pm$ standard deviation. For both approaches, AVC improves output
  quality across all nine metrics.}
  \label{tab:bicubic-baseline}
  \resizebox{\textwidth}{!}{%
  \begin{tabular}{@{}l|cc|cc@{}}
    \toprule
    & \multicolumn{2}{c|}{Proposed system} & \multicolumn{2}{c}{Bicubic interpolation} \\
    \cmidrule(lr){2-3} \cmidrule(lr){4-5}
    Metric & Without AVC & With AVC & Without AVC & With AVC \\
    \midrule
    MAE [$\downarrow$]
      & 0.036 $\pm$ 3.05e-3 & 0.011 $\pm$ 5.89e-4
      & 0.037 $\pm$ 3.04e-3 & 0.016 $\pm$ 6.22e-4 \\
    RMSE [$\downarrow$]
      & 0.091 $\pm$ 6.79e-3 & 0.028 $\pm$ 7.69e-3
      & 0.092 $\pm$ 6.61e-3 & 0.035 $\pm$ 6.15e-3 \\
    PSNR [$\uparrow$]
      & 20.8 $\pm$ 6.50e-1 & 31.5 $\pm$ 2.60
      & 20.707 $\pm$ 6.26e-1 & 29.313 $\pm$ 1.58 \\
    SSIM [$\uparrow$]
      & 0.849 $\pm$ 2.40e-2 & 0.972 $\pm$ 3.12e-3
      & 0.831 $\pm$ 2.40e-2 & 0.938 $\pm$ 3.54e-3 \\
    ERQA [$\uparrow$]
      & 0.434 $\pm$ 7.43e-2 & 0.842 $\pm$ 1.23e-2
      & 0.461 $\pm$ 6.74e-2 & 0.756 $\pm$ 8.06e-3 \\
    MS-SSIM [$\uparrow$]
      & 0.927 $\pm$ 2.00e-2 & 0.990 $\pm$ 7.63e-4
      & 0.920 $\pm$ 2.03e-2 & 0.976 $\pm$ 1.16e-3 \\
    DISTS [$\downarrow$]
      & 0.196 $\pm$ 1.89e-2 & 0.087 $\pm$ 1.38e-2
      & 0.188 $\pm$ 2.07e-2 & 0.075 $\pm$ 1.09e-2 \\
    FSIM [$\uparrow$]
      & 0.894 $\pm$ 2.21e-2 & 0.990 $\pm$ 1.15e-3
      & 0.884 $\pm$ 2.22e-2 & 0.973 $\pm$ 1.62e-3 \\
    PieAPP [$\downarrow$]
      & 3.009 $\pm$ 2.75e-1 & 0.612 $\pm$ 5.88e-2
      & 3.067 $\pm$ 2.40e-1 & 0.714 $\pm$ 5.24e-2 \\
    \bottomrule
  \end{tabular}}
\end{table}

\section{Real-world Applicability and Limitations}
Our experiments indicate that the processing of an SR burst of 32 images (16 image pairs) with a resolution of $4096 \times 2160$ (output resolution of $8192 \times 4320$) requires 17 seconds on an office workstation (CPU: Intel Core i9-12900H, GPU: NVIDIA GeForce RTX 3080) and roughly triple that on a Jetson-class embedded computer.
At an operating altitude of $h=20$ km, using the presented cameras with a $16\degree$ camera field-of-view, the ground footprint in the imaging direction is $W\approx 5.62$ km. A platform traveling at $v=10$ m/s moves $\Delta=510$ m during an SR processing time of $t=51$ s. The along-track overlap between consecutive bursts is therefore $(W-\Delta)/W \approx  0.9$, i.e., about $90\%$, meaning that the scene overlap available for multi-frame SR remains very high, supporting near-real-time continuous evaluation under the stated assumptions. Note that the speed used for the approximate calculation corresponds to the maximum speed reported for the Zephyr aircraft \cite{Zephyr_weaver2019}.

While no aerodynamic loading or atmospheric effects were introduced on the platform at this stage, the resulting vibration spectra and image sequences already provide a useful reference for how the system behaves before it is exposed to the more complicated aeroelastic phenomena expected in high-altitude operation.

Further testing and refinement are required to ensure the suitability of the presented framework for operation in real stratospheric environments. Additional factors, such as gust-induced dynamics, temperature-dependent sensor behavior, and optical disturbances (including haze, refraction, and varying solar illumination) are expected to increase the difficulty of image registration and reconstruction. These currently unmodeled real-world effects are also likely to increase uncertainty and parameter variability, underscoring the need to extend the current nominal $\mathcal{H}_\infty$ controller toward a more robust control synthesis framework in future work.

\section{Conclusion}
This paper presents a methodology that extends the knowledge on advancements in Earth Observation from high-altitude remote sensing platforms.
Given the limited payload capacity and the flexible, high-aspect-ratio wings of such platforms, our proposed distributed camera system enhances imaging quality through software-based post-processing together with AVC, as an alternative to expensive or heavy optical systems.
To validate the concept, real-time tests were conducted on a dedicated experimental test bench with two lightweight cameras.
The results show that the designed controller effectively suppressed vibrations, improving the input images for the SR framework and thereby significantly enhancing the final image quality according to the defined metrics.

The demonstrator platform's architecture and the SR pipeline have the ability to support more than two sensors. Future research and development will focus on integrating diverse sensor modalities, such as multispectral cameras, to enrich the captured information and improve robustness. Additionally, since several real-world atmospheric effects have not yet been considered (e.g., long-range haze, significant illumination variations), evaluating the system under properly simulated atmospheric and environmental conditions will be essential for assessing its reliability and generalization across different operational scenarios.

\section*{Acknowledgment}
The research reported in this paper is part of
project no. TKP-NVA-01, implemented with the support provided by the Ministry of Innovation and Technology of Hungary from the National Research, Development and Innovation Fund, financed under the TKP2021 funding
scheme.
Cooperative Technologies National Laboratory, Project no. 2022-2.1.1-NL-2022-00012 has been implemented with the support provided by the Ministry of Culture and Innovation of Hungary from the National Research, Development and Innovation Fund, financed under the National Laboratories funding scheme.
The work presented in this paper received funding from the ELKH Pseudo Satellite Research Grant and from the Hungarian Scientific Research Fund (No. NKFIH OTKA K-139485).

The authors would like to thank Marcell Golarits for his contributions to setting up the testing environment and for the insightful discussions that supported this work.

\bibliographystyle{IEEEtran}
\bibliography{references}

\end{document}